\documentclass{pasj00}
\begin{document}
\SetRunningHead{Y. Takeda and M. Takada-Hidai}{Carbon Abundances of Metal-Poor 
Stars from C~I 1.068--1.069~$\mu$m Lines}
\Received{2013/01/07}%{yyyy/mm/dd}
\Accepted{2013/01/26}%{yyyy/mm/dd}

\title{Carbon Abundances of Metal-Poor Stars\\
Determined from the C~I 1.068--1.069~$\mu$m Lines
\thanks{Based on data collected at Subaru Telescope, which is operated by 
the National Astronomical Observatory of Japan.}
\thanks{The large data tables are separately provided in the machine-readable 
form as electronic tables E1 and E2.}
}

%%% Please use the following style in case that sorting by 
%%% affiliation is impossible. 
%
% \author{%
%   D-Firstname \textsc{D-Familyname}\altaffilmark{1}
%   E-Firstname \textsc{E-Familyname}\altaffilmark{1,2}
%   and
%   F-Firstname \textsc{F-Familyname}\altaffilmark{2}}
% \altaffiltext{1}{Address of Institute}
% \email{ddddd@xxx.xxx.xx.xx}
% \email{eeeee@xxx.xxx.xx.xx}
% \altaffiltext{2}{Address of Institute}

\author{Yoichi \textsc{Takeda}}
\affil{National Astronomical Observatory of Japan 
2-21-1 Osawa, Mitaka, Tokyo 181-8588}
\email{takeda.yoichi@nao.ac.jp}
\and 
\author{Masahide \textsc{Takada-Hidai}}
\affil{Liberal Arts Education Center, Tokai University, 
4-1-1 Kitakaname, Hiratsuka, Kanagawa 259-1292}
\email{mth\_tsc@tsc.u-tokai.ac.jp}

%%% Please use the following style in case that sorting by 
%%% affiliation is impossible. 
%
% \author{%
%   D-Firstname \textsc{D-Familyname}\altaffilmark{1}
%   E-Firstname \textsc{E-Familyname}\altaffilmark{1,2}
%   and
%   F-Firstname \textsc{F-Familyname}\altaffilmark{2}}
% \altaffiltext{1}{Address of Institute}
% \email{ddddd@xxx.xxx.xx.xx}
% \email{eeeee@xxx.xxx.xx.xx}
% \altaffiltext{2}{Address of Institute}

%% `\KeyWords{}' always has to be placed before `\maketitle'.
%\KeyWords{xxxx:xxxx ......} %Do NOT move this preamble from here!

\maketitle

\begin{abstract}

A non-LTE analysis of C~{\sc i} lines at 1.068--1.069~$\mu$m 
was carried out for selected 46 halo/disk stars covering 
a wide metallicity range ($ -3.7\ltsim$~[Fe/H]~$\ltsim +0.3$), 
based on the spectral data collected with IRCS+AO188 of 
the Subaru Telescope, in order to examine whether and how 
these strong neutral carbon lines of multiplet 1 can be exploited 
for establishing stellar carbon abundances, especially for very 
metal-poor stars where CH molecular lines have been commonly used. 
These C~{\sc i} lines were confirmed to be clearly visible for 
all stars down to [Fe/H]~$\sim -3.7$, from which C abundances 
could be successfully determined. The resulting [C/Fe] vs. [Fe/H]
diagram revealed almost the same trend established from previous studies.
When the results for individual stars are compared with the 
published data collected from various literature, while a reasonable 
agreement is seen as a whole, a tendency is observed that 
our abundances are appreciably higher than those from CH lines
especially for very metal-poor giants of low gravity.
Since the abundances of these C~{\sc i} lines are subject to rather 
large non-LTE corrections (typically by several tenths dex) 
whose importance progressively grows as the metallicity is lowered, 
attention should be paid to how the collisional rates (especially 
due to neutral hydrogen) are treated in non-LTE calculations. 

\end{abstract}

%\section{}
%
%\noindent IMPORTANT NOTICE\\
%1. ``\verb|\draft|'' creates single column and double spaces format.\\
%2. If you comment out ``\verb|\draft|'', the output will be double column
%   and single space.\\
%3. For cross-references, the use of ``\verb|\label|, \verb|\ref|, \verb|\cite|'%' 
%   and the thebibliography environment is strongly recommended. \\
%4. Do NOT use ``\verb|\def|, \verb|\renewcommand|''.\\
%5. Do NOT redefine commands provided by PASJ00.cls.\\

%\newpage

%Sect. 1
\section{Introduction}

The abundances of carbon (one of the most abundant elements
next to hydrogen, helium, and oxygen) play a significant role 
in stellar spectroscopy. Above all, the behavior of C in 
the very metal-poor regime ([Fe/H]~$\ltsim -2$) is important,
since various peculiarities (i.e., enhancement as well as 
deficiency) are observed, which have attracted interest of 
astrophysicists in the field of galactic nucleosynthesis or 
stellar evolution. Thus a number of carbon abundance studies
of metal-deficient stars have been published so far, as compiled 
in SAGA\footnote{Available on-line at 
$\langle$http://saga.sci.hokudai.ac.jp/$\rangle$.}
 (Stellar Abundances for the Galactic Archeology) database 
(Suda et al. 2008, 2011). Regarding the spectral lines
to be invoked for spectroscopically determining the abundance 
of carbon, several indicators are known to usable, such as atomic 
lines (C~{\sc i}) or molecular lines (CH, C$_{2}$), each of which 
have different visibilities depending on the type of stars,
and have different characteristics; e.g., in terms of 
the sensitivity to the 3D effect or non-LTE effect 
(cf. Asplund 2005; subsection 3.3 therein).

Focusing on very metal-deficient stars in the galactic halo
($-4 \ltsim$~[Fe/H]~$\ltsim -2$), we note that most
studies on carbon abundances have been done based on CH molecular lines
in the blue region (G band) which are so strong as to be visible
even at such an extremely low metallicity. Although 
a group of neutral carbon lines at 9061--9111~$\rm\AA$ 
(multiplet 3; 3s~$^{3}{\rm P}^{\rm o}$ -- 3p $^{3}{\rm P}$;
$\chi_{\rm low} \sim 7.5$~eV) were sometimes used for metal-poor dwarfs
by several investigators (Tomkin et al. 1992; Akerman et al. 2004;
Takeda \& Honda 2005; Fabbian et al. 2006, 2009), they have not been
the mainstream (presumably because they may suffer an appreciable 
non-LTE effect).

Here, we should realize that other C~{\sc i} lines are available
in near-IR region, which are namely the lines at $\sim$~1.07~$\mu$m
(multiplet 1; 3s~$^{3}{\rm P}^{\rm o}$ -- 3p $^{3}{\rm D}$;
$\chi_{\rm low} \sim 7.5$~eV) originating from the same lower term
(3s~$^{3}{\rm P}^{\rm o}$) as multiplet 3 lines. Actually, they are 
even more suitable than C~{\sc i} 0.91~$\mu$m lines, because of 
their strengths (the transition probability of the strongest 
C~{\sc i} 10691 line in multiplet 1 is larger by 0.2~dex than 
the strongest C~{\sc i} 9094.83 line in multiplet 3; cf. table 2) 
as well as their location in the spectral region almost 
free from telluric lines (these are
actually a nuisance problem for C~{\sc i} 0.91~$\mu$m lines).
However, to our knowledge, these lines have barely been used for
carbon abundance determinations, except for the Sun and Vega
(e,g., St\"{u}renburg \& Holweger 1990; Takeda 1992, 1994).
Why not exploit these lines to study carbon abundances of
metal-poor stars by making use of these merit?

Timely, we recently obtained near-IR ($zJ$-band; 1.04--1.19~$\mu$m) 
high-dispersion spectra of 46 disk/halo stars (dwarfs as well giants) 
in a wide metallicity range ($-3.7 \ltsim$~[Fe/H]~$\ltsim +0.3$),
by using the Subaru Telescope with IRCS+AO188, 
and carried out sulfur abundance determinations based on
the S~{\sc i} triplet lines at 10455--10459~$\rm\AA$
(Takeda \& Takada-Hidai 2011a, 2012; hereinafter referred to as 
Paper I and Paper II, respectively).
We thus decided to study the strong C~{\sc i} lines of multiplet 1 
at $\sim$~1.07~$\mu$m in these spectra, in order see whether the C abundances 
of very metal-poor stars can be successfully determined by these 
lines and how they are compared with the published results derived
from other lines (such as those of CH). This is the purpose of 
this investigation.

The remainder of this article is organized as follows. After 
describing the observational data in section 2, we explain 
the details of abundance determinations (stellar parameters, 
treatment of non-LTE effect, method of analysis, uncertainties, 
etc.) in section 3. Section 4 is devoted to the discussion section,
where the results are examined, compared with the published work,
and the merits and problems of these near-IR C~{\sc i} lines are 
discussed. The summary is given in section 5. In the Appendix, 
the grid of theoretical non-LTE corrections for the C~{\sc i} 9061--9111 
lines of multiplet 3 are re-presented, since a mis-treatment was 
found in our previously published results. 

%Sect. 2 (figure 1, table 1)
\section{Observational Data}

The observational data used for this study are
the high-dispersion ($R\sim 20000$) $zJ$-band (1.04--1.19~$\mu$m)
spectra obtained by using IRCS+AO188 of the Subaru Telescope 
in two observing periods of 2009 July 29 and 30 (UT) as well as 
2011 August 17 and 18 (UT). In the former 2009 July run, 33 comparatively 
bright stars of wide varieties (from [Fe/H]~$\sim -3.7$ to $\sim +0.3$, 
dwarfs as well as giants) were observed,\footnote{In addition, Altair
(A7~V, rapid rotator with $v_{\rm e}\sin i \sim 200$~km~s$^{-1}$)
was also observed to be used as a reference of telluric lines.} 
while selected 13 very metal-poor stars ([Fe/H]~$\ltsim -2$) and Vesta 
(substitute for the Sun) were targeted in the latter 2011 August run. 
This makes 47 ($=33+13+1$) spectra as a total, though the net number
is 46 because G~64-37 was repeatedly observed.
See section 2 of Papers I and II for more details of the observations
and the data reduction. The list of these 47 objects is given
in table 1, where the targets in 2009 and those in 2011 are
separately presented. 

The 2009 July spectra for representative 4 stars of different 
metallicities along with Altair in the 10650--10750~$\rm\AA$ region 
are displayed in figure 1. 
Note that most of the conspicuous lines are those of neutral
carbon and silicon, and several strong C~{\sc i} lines
(e.g., at 10683, 10691, 10730~$\rm\AA$) are clearly visible
also for the most metal-poor star (BD~+44~493).
Besides, we can recognize from this figure (see the broad-line 
spectrum of Altair) that this region is practically free from 
any telluric lines, in contrast to the C~{\sc i} 9061--9111
region influenced by H$_{2}$O lines of earth's atmosphere origin.

%Figure 1
\setcounter{figure}{0}
\begin{figure}
  \begin{center}
    \FigureFile(70mm,70mm){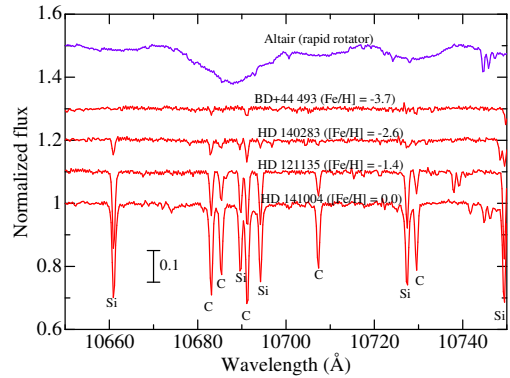}
    %%% \FigureFile(width,height){filename}
  \end{center}
\caption{Spectra of representative 4 stars with different metallicities
in the 10650--10750~$\rm\AA$ region comprising C~{\sc i} and Si~{\sc i}
lines, where the wavelength scale of stellar lines is adjusted to 
the laboratory frame. The spectrum of Altair is also shown to 
demonstrate the negligible effect of telluric lines.
}
\end{figure}

%Sect. 3 (figure 2, figure 3, figure 4, figure 5, table 2)
\section{Abundance Determination}

\subsection{Non-LTE Calculations and Model Atmospheres}

Regarding the atmospheric parameters ($T_{\rm eff}$, $\log g$, 
$v_{\rm t}$, and [Fe/H]) of the program stars necessary for 
constructing model atmospheres and determining abundances, 
we used exactly the same values as adopted in Papers I and II,
in which various published studies were consulted (cf. caption
in table 1 of Papers I and II for the individual reference 
sources). These parameters are presented in table 1.

As to the computation of non-LTE departure coefficients,
we closely followed our previous work (see Takeda 1992, 1994; 
Takeda \& Honda 2005 for the details). Practically, our
non-LTE calculations were carried out on a large grid of 150 
($= 6 \times 5\times 5$) models corresponding to combinations of 
six $T_{\rm eff}$ (4500, 5000, 5500, 6000, 6500, 7000~K),
five $\log g$ (1, 2, 3, 4, 5), and five [Fe/H]
($-4$, $-3$, $-2$, $-1$, 0), where [C/Fe]~=~0 (carbon abundance) 
and $v_{\rm t}$ = 2~km~s$^{-1}$ (microturbulence) were assumed. 

To keep consistency with our previous studies, the collsional 
rates ($C_{\rm e}$ for electron collisions and $C_{\rm H}$ for
neutral hydrogen collisions) were computed by following the 
recipe described in subsubsection 3.1.3 of Takeda (1991).
Namely, regarding electron collisions, we invoked 
Van Regemorter's (1962) formula for permitted transitions 
($C_{\rm e}^{\rm (p)}$) and Auer and Mihalas's (1973) equation (13)
for forbidden transitions ($C_{\rm e}^{\rm (f)}$). 
Meanwhile, as for hydrogen collision rates, $C_{\rm H}^{\rm (p)}$ 
for permitted transition was evaluated by invoking 
Steenbock and Holweger's (1984) formula, which is based on 
Drawin's (1968, 1969) classical cross section, and $C_{\rm H}^{\rm (f)}$
for forbidden transitions was approximated by simply scaling
$C_{\rm e}^{\rm (f)}$ according to the collision frequency
(determined by the relative speed and the density of colliding 
particles) on the assumption of the same cross section.
Since such computed 
$C_{\rm H} (= C_{\rm H}^{\rm (p)} + C_{\rm H}^{\rm (f)})$ has
only order-of-magnitude accuracy at best and tends to be
often overestimated, we also tested an alternative case of 
reduced $C_{\rm H}$ by introducing a reduction factor $k$ 
as was done in Takeda and Honda (2005; cf. appendix 1 therein):
\begin{equation}
C_{\rm total} = C_{\rm e} + k C_{\rm H},
\end{equation}
though we still assume $k=1$ as the standard case to be adopted.
Such a scaling of $C_{\rm H}$ with $k$ is applied also to bound-free 
collisional ionization rates, for which we invoked Jefferies (1968) 
for electron collisions and Steenbock and Holweger (1984) 
for neutral-hydrogen collisions.

We then interpolated Kurucz's (1993) grid of ATLAS9 model 
atmospheres as well as the grid of the non-LTE departure 
coefficients in terms of $T_{\rm eff}$, $\log g$, and [Fe/H] to 
generate the atmospheric model and the departure coefficient data 
for each star. 

Besides, for the reader's convenience, extensive grids of 
non-LTE as well as LTE equivalent widths ($EW^{\rm N}$ and $EW^{\rm L}$) 
and non-LTE abundance corrections ($\Delta$) for five multiplet 1 
lines (C~{\sc i} 10683, 10685, 10691, 10707, 10729) were further 
computed for three [C/Fe] values ($-0.3$, 0.0, and $+0.3$) 
and three $v_{\rm t}$ values (1, 2, and 3~km~s$^{-1}$), 
based on the non-LTE departure coefficients computed for 150 model 
atmospheres with two different treatments of neutral hydrogen 
collisions ($k=1$ and $k=0.1$), which are presented in electronic table E1.

\subsection{Synthetic Spectrum Fitting}

Inspecting the spectrum feature around $\sim$~1.07~$\mu$m (cf. 
figure 1), we decided to concentrate on the 10680--10697~$\rm\AA$
region which comprises three C~{\sc i} lines (in which C~{\sc i}~10691 
is the strongest one of multiplet 1) and two Si~{\sc i} lines.
Then, following the similar way as in Papers I and II, we carried out 
non-LTE spectrum-synthesis analyses by applying Takeda's (1995) 
automatic fitting procedure to this 17~$\rm\AA$-interval region 
while regarding the non-LTE carbon abundance ($A_{\rm C}^{\rm N}$),
LTE silicon abundance ($A_{\rm Si}^{\rm L}$), 
the macro-broadening parameter ($v_{\rm M}$; $e$-folding half-width
of the Gaussian macrobroadening function; 
$\propto \exp[-(v/v_{\rm M})^2]$), and the wavelength 
shift ($\Delta \lambda$) as adjustable free parameters 
to be varied until a convergence is accomplished.
The adopted atomic data of the relevant C~{\sc i} and Si~{\sc i} lines 
are presented in table 2.
This fitting procedure turned out quite successful and
$A_{\rm C}^{\rm N}$ could be established for all 47 cases,
though we had to fix the silicon abundance for two extremely 
metal-poor stars (BD~+44~493 and G~64-37 [2009 July data]) 
where Si~{\sc i} lines are hardly visible.
How the theoretical spectrum for the converged solutions 
fits well with the observed spectrum is displayed in 
figure 2 (2009 July data) as well as figure 3 (2011 August data).
(Note that these two figures are arranged in analogy with
figure 2 of Paper I and figure 1 of Paper II, respectively.)
The resulting non-LTE C abundances ($A_{\rm C}^{\rm N}$) are 
given in table 1, where the values of [C/Fe]
($\equiv A_{\rm C}^{\rm N} - A_{\rm C \odot}^{\rm N}$)~$-$~[Fe/H];
 with $A^{\rm N}_{\odot}$ = 8.54 from Vesta) are also presented.
More detailed results are found in electronic table E2, 
where the results of $A_{\rm Si}^{\rm L}$ and $v_{\rm M}$ are
also given.

\subsection{Abundance-Related Quantities}

While the non-LTE synthetic spectrum fitting directly yields
the final abundance solution, this approach is not necessarily
suitable when one wants to evaluate the extent of non-LTE 
corrections or to study the abundance sensitivity to changing 
the atmospheric parameters (i.e., it is tedious to repeat 
the fitting process again and again for different assumptions 
or different atmospheric parameters).
Therefore, with the help of Kurucz's (1993) WIDTH9 program 
(which had been considerably modified in various respects; 
e.g., inclusion of non-LTE effects, etc.), we computed 
the equivalent widths for each of the three C~{\sc i} lines 
($EW_{10683}$, $EW_{10685}$, and $EW_{10691}$)  ``inversely'' 
from the abundance solution (resulting from non-LTE spectrum 
synthesis) along with the adopted atmospheric model/parameters, 
since they are much easier to handle.
Based on such evaluated $EW$ values, the LTE abundances for each of
the lines ($A^{\rm L}_{10683}$, $A^{\rm L}_{10685}$, and 
$A^{\rm L}_{10691}$) were freshly computed, from which the 
non-LTE corrections ($\Delta_{10683}$, $\Delta_{10685}$, and 
$\Delta_{10691}$) were derived such as 
$\Delta_{10683} \equiv A^{\rm L}_{10683} - A_{\rm C}^{\rm N}$, etc.

Besides, in order to examine the effect of changing $C_{\rm H}$ 
(neutral hydrogen collision rates), we also computed 
$\Delta_{10691}$ ($k=0.1$) from $EW_{10691}$
by using the non-LTE departure coefficients calculated for the $k=0.1$ 
case (i.e., $C_{\rm H}$ is reduced by 1/10 from the adopted standard
case of $k=1$) and evaluated the difference
\begin{equation}
\delta\Delta_{10691} \equiv \Delta_{10691} (k=0.1) - \Delta_{10691} (k=1).
\end{equation}

We then estimated the uncertainties in $A_{\rm C}^{\rm N}$
by repeating the analysis on $EW_{10691}$ ($EW$ for the strongest
component) while perturbing the standard values of atmospheric 
parameters interchangeably by $\pm 100$~K in $T_{\rm eff}$, 
$\pm 0.2$~dex in $\log g$, and $\pm 0.3$~km~s$^{-1}$ in 
$v_{\rm t}$ (which we regarded as typical uncertainties 
of the atmospheric parameters according to the original references;
cf. subsection 4.2 in Paper I). 
Let us call these six kinds of abundance variations as
$\delta_{T+}$, $\delta_{T-}$, $\delta_{g+}$, $\delta_{g-}$, 
$\delta_{v+}$, and $\delta_{v-}$, respectively.
We then computed the root-sum-square of three quantities
$\delta_{Tgv} \equiv (\delta_{T}^{2} + \delta_{g}^{2} + \delta_{v}^{2})^{1/2}$
as the abundance uncertainty (due to combined errors in 
$T_{\rm eff}$, $\log g$, and $v_{\rm t}$), 
where $\delta_{T}$, $\delta_{g}$, and $\delta_{v}$ are defined as
$\delta_{T} \equiv (|\delta_{T+}| + |\delta_{T-}|)/2$, 
$\delta_{g} \equiv (|\delta_{g+}| + |\delta_{g-}|)/2$, 
and $\delta_{v} \equiv (|\delta_{v+}| + |\delta_{v-}|)/2$,
respectively. 

The resulting [C/Fe], $\Delta_{10691}$, $\delta \Delta_{10691}$, 
and $EW_{10691}$ for each star are plotted against $T_{\rm eff}$ 
in figures 4 a--d, where $\log g > 3.0$ stars (dwarfs) 
and $\log g < 3.0$ stars (giants) are distinguished by filled and 
open symbols, respectively, and the error bar attached to [C/Fe] 
represents $\pm \delta_{Tgv}$. 
While only $A_{\rm C}^{\rm N}$, $EW_{10691}$, $\Delta_{10691}$, 
and [C/Fe] are given in table 1, all the relevant 
data (including $EW$ and $\Delta$ for all three lines and each of 
the $\delta$ values) are presented in electronic table E2.

%Figure 4
\setcounter{figure}{3}
\begin{figure}
  \begin{center}
    \FigureFile(70mm,100mm){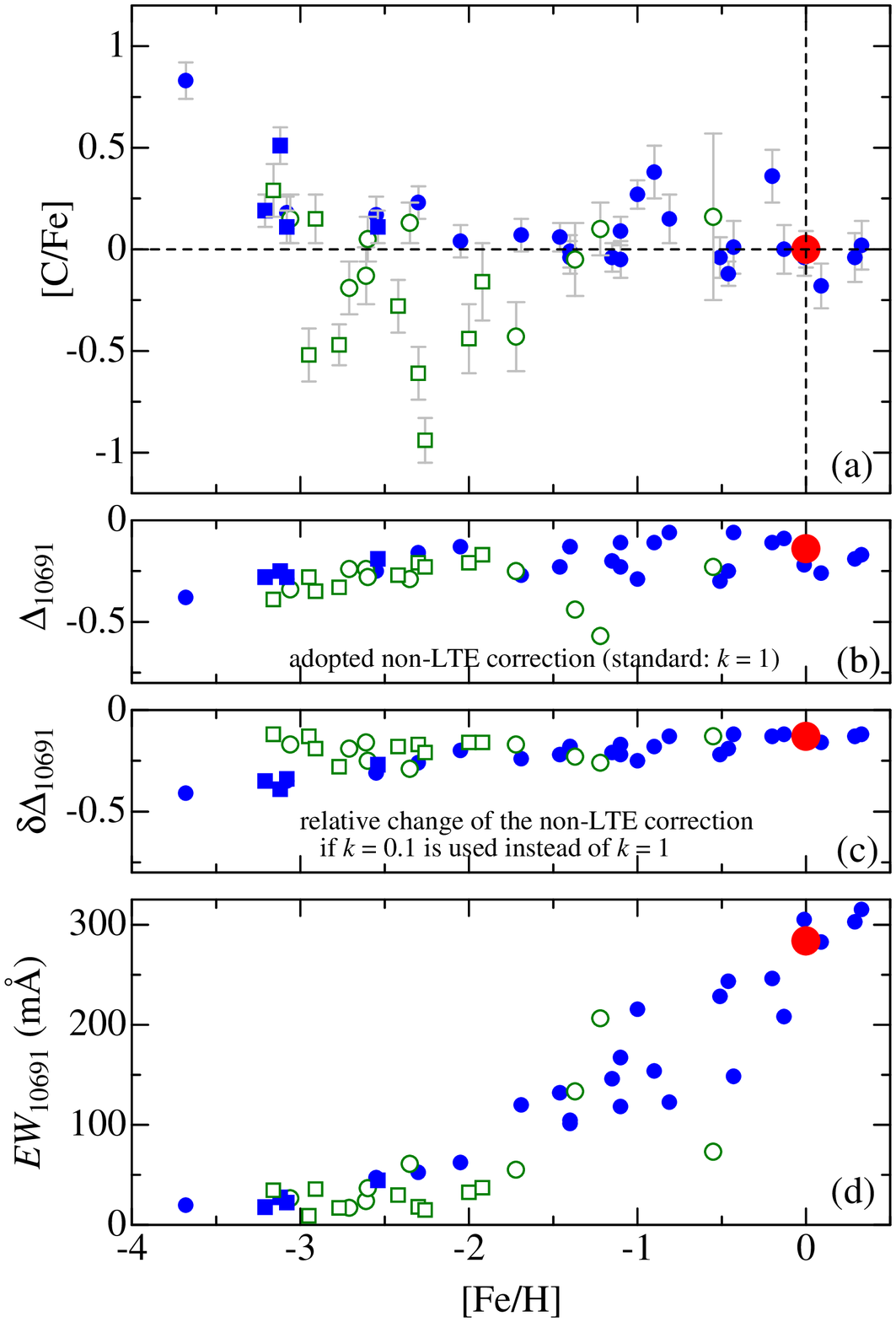}
    %%% \FigureFile(width,height){filename}
  \end{center}
\caption{
Carbon abundances (along with the related quantities) plotted against [Fe/H]: 
(a) [C/Fe] (C-to-Fe logarithmic abundance ratio corresponding to non-LTE 
carbon abundance; where attached error bars represent the ambiguities
due to uncertainties in the atmospheric parameters ($\delta_{Tgv}$;
cf. subsection 3.3)
(b) $\Delta_{10691}$ (non-LTE correction for the C~{\sc i}~10691 line
derived with the standard classical treatment of neutral-hydrogen collisions
with $k=1$).
(c) $\delta \Delta_{10691}$ (relative difference of non-LTE correction 
caused by reducing the neutral-hydrogen collisions by 1/10; 
cf. equation (2)). 
(d) $EW_{10691}$ (equivalent width for the C~{\sc i}~10691 line).
Dwarfs ($\log g > 3$) and giants ($\log g < 3$) are discriminated by filled 
(blue) and open (green) symbols, respectively. The results for 33 stars 
based on the 2009 July data and those for 12 stars based on the 2011 August 
data are shown by circles and squares, respectively.
The large red filled circle denotes the Sun (Vesta). 
}
\end{figure}

It can be seen from figure 4a that the typical extent of $\delta_{Tgv}$
is $\ltsim$~0.1--0.2~dex and not very significant.
We should bear in mind, however, that the non-LTE corrections are so 
large and that some uncertainties in them are more or less inevitable. 
Apart from the ambiguities in the treatment of collisional
rates due to neutral hydrogen collisions (choice of $k$)
as discussed later in subsections 4.1 and 4.3, we should recall
that our non-LTE calculations were done on the condition of [C/Fe] = 0.
Actually, this assumption is reasonable, since most of the [C/Fe] 
values derived for our targets scatter around $\sim 0$ with a dispersion 
of $\sim \pm 0.5$~dex (figure 4a). However, [C/Fe] values of two stars
considerably depart from zero; i.e., +0.83 (BD~+44~493) and 
$-0.94$ (HD~13979), for which the mismatch of [C/Fe] may cause 
some errors. As a test, we evaluated the abundance change
when we used departure coefficients computed with [C/Fe] = +1.0
(BD~+44~493) and [C/Fe] = $-1.0$ (HD~13979) instead of [C/Fe] = 0.
These calculations were done on the selected two models 
(t55g40m40 and t50g20m20 included in the model grid of electronic 
table E1) having parameters closest to those of these two stars. 
We then found that the extent of non-LTE overpopulation for the relevant 
high-excitation term is enhanced/reduced by decreasing/increasing 
the C abundance,\footnote{This trend could be interpreted as due to 
the efficiency change in the UV pumping from ground or low-excitation 
levels to upper levels of high excitation, which may possibly depend 
upon the C abundance in the very metal-poor regime. That is, as long as 
such UV transitions are optically thick and in radiative detailed 
balance, such process does not occur. Meanwhile, as the C abundance 
is sufficiently lowered and some of these UV transitions eventually 
become optically thin, high-excitation C~{\sc i} levels may get 
overpupulated via upward pumping by absorbing hot UV continuum radiation.} 
and thus the $A_{\rm C}^{\rm N}$ (derived for the same $EW$) is 
increased by 0.25~dex ([C/Fe]~=~+1.0; BD~+44~493) 
and reduced by 0.10 dex ([C/Fe]~=~$-1.0$; HD~13979) compared to
the results for the standard [C/Fe]~=~0.0 case.
Accordingly, the error caused by applying the departure coefficients 
computed with [C/Fe]~=~0 to such exceptional cases of considerably
peculiar C abundance would be $\sim$~0.1--0.2~dex.

%Sect. 4 (figure 5, figure 6, table 3)
\section{Discussion}

\subsection{Line Strength and Non-LTE Effect}

According to figures 4b--d and table 1, the following features
are recognized regarding the behavior of the strength and
the non-LTE correction of the C~{\sc i} 10691 line (the strongest 
among the multiplet 1 lines at $\lambda$~1.07~$\mu$m).\\
--- The C~{\sc i} 10691 line is clearly visible even for extremely 
metal-poor stars ([Fe/H]~$\ltsim -3$) with $EW$ on the order of 
$\sim 10$~m$\rm\AA$, irrespective of dwarfs or giants, 
which suggests the usability of this line to explore the carbon 
synthesis history of the Galaxy.\\
--- The non-LTE correction is always negative (corresponding
to non-LTE line-strengthening). While the typical extent of 
$|\Delta_{10691}|$ for the standard $k=1$ case is several tenths dex 
($\ltsim 0.5$~dex; cf. figure 4b), it is further raised by 
almost a similar amount when $k$ is reduced from 1 to 0.1 
(figure 4d) which makes $|\Delta_{10691} (k=0.1)|$ even up to 
$\ltsim 1$~dex. This implies that the non-LTE correction is
sensitive to how the neutral hydrogen collision is treated.\footnote{
This does not agree with the conclusion of Fabbian et al. 
(2006, 2009), who found that the non-LTE corrections for
C~{\sc i} lines are insensitive to a choice of the reduction factor 
($S_{\rm H}$ in their notation); even changing $S_{\rm H} =1$ to 
$S_{\rm H} =0$ makes only an insignificant change ($\sim 0.1$~dex) 
in their non-LTE corrections for C~{\sc i} 9061--9111 lines of 
multiplet 3, which are as large as those of C~{\sc i}~10683--10691 lines
of multiplet 1 (cf. figure 8b). Considering the more or less reasonable 
consistency between the $\Delta$ (multiplet 3) values of Takeda and Honda 
(2005) (see also the Appendix of this paper) and those of Fabbian et al. 
(2006; cf. their figure 8), this discrepancy is hard to understand. 
Although the cause is not clear, something may be different 
between their and our calculation procedures.}\\
--- The extent of $|\Delta_{10691}|$ systematically increases with 
a decrease in [Fe/H], despite that the line strength is weakened, 
which means that the non-LTE correction becomes progressively
more important as we go toward the extremely metal-poor regime.

%Figure 5
\setcounter{figure}{4}
\begin{figure}
  \begin{center}
    \FigureFile(70mm,70mm){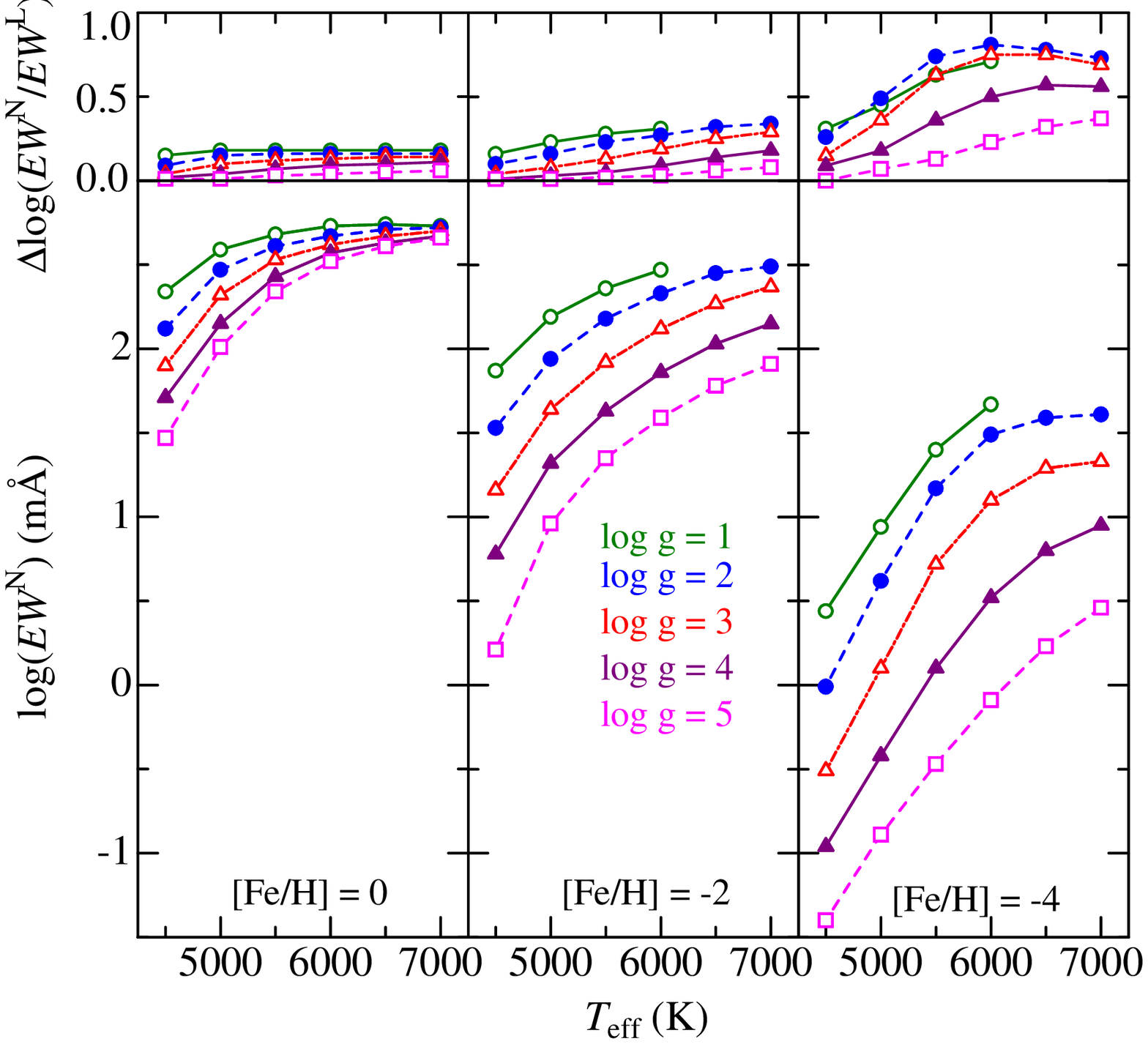}
    %%% \FigureFile(width,height){filename}
  \end{center}
\caption{Theoretically computed non-LTE logarithmic equivalent widths
($\log EW^{\rm N}$; lower panels) and the difference between non-LTE 
and LTE logarithmic equivalent widths ($\log EW^{\rm N} - \log EW^{\rm L}$;
upper panels) plotted against $T_{\rm eff}$, which were calculated
for the C~{\sc i} 10691.24 line
with the standard treatment of H~{\sc i} collision cross sections
($k = 1$), $v_{\rm t}$ = 2~km~s$^{-1}$, and [C/Fe] = 0. 
The left, middle, and right panels correspond to [Fe/H] = 0, 
$-2$, and $-4$, respectively. Open circles, filled circles,
open triangles, filled triangles, and open squares denote
results for $\log g$ = 1, 2, 3, 4, and 5, respectively.
}
\end{figure}

In order to help our understanding of these features mentioned above,
we display in figure 5 how $EW^{\rm N}$ (theoretical non-LTE equivalent 
width computed for the C~{\sc i} 10691 line) and $EW^{\rm N}/EW^{\rm L}$
(non-LTE to LTE equivalent-width ratio) depend on $T_{\rm eff}$,
$\log g$, and [Fe/H], based on the data in electronic table E1 (cf.
subsection 3.1) corresponding to the standard $k=1$ case.
We can learn from this figure significant characteristics
concerning the line-strength behaviors:\\
--- While it is natural that the line is weakened with a
decrease in [Fe/H] (=[C/H]), its strength grows with an increase 
in $T_{\rm eff}$ as well as with a decrease in $\log g$.\\
--- The non-LTE to LTE strength ratio is always larger than 
unity (i.e., non-LTE line strengthening) as we already recognized
from the negative sign of $\Delta$, which increases with a 
lowering of the metallicity; it amounts up to $\gtsim 5$ at [Fe/H] = 
$-4$, which explains the [Fe/H]-dependence of $|\Delta|$ mentioned
above.\\
--- We may interpret these trends in terms of the ionization degree
of carbon (partially ionized in the atmospheres of FGK stars), 
which tends to enhance with an increase in $T_{\rm eff}$
as well as with a decrease in $\log g$ (lowered density) and [Fe/H]
(lowered electron density, enhanced UV overionizing radiation 
due to the increased transparency), while the change in the H$^{-}$ 
continuum opacity is also involved especially in the $\log g$-effect.
That is, the population of high-excitation C~{\sc i} lines is
closely coupled with the population of C~{\sc ii} parent term
(i.e., the effect of $n^{\rm (II)}/n^{\rm (I)\dagger}$; cf. section 3
in Takeda 1992), which makes the population of high-excitation 
C~{\sc i} level larger as the ionization fraction 
(C~{\sc ii}/C~{\sc i}) increases. Therefore, the degree of ionization
plays an important role in understanding the strengths and the non-LTE
effect of high-excited C~{\sc i} lines such as those of
multiplet 1 or 3.\\
--- Besides, under the considerably metal-poor condition of very low
C abundance, some of the strong C~{\sc i} transitions in UV would 
become optically thin, which may act to enhance non-LTE overpopulation 
of high-excitation levels via the pumping effect (cf. footnote 3).\\
--- Thus, the mechanism of non-LTE line intensification is not 
necessarily the same for the strong-line case of near-solar metallicity 
and the weak-line case of very low metallicity. In the former situation,
the line is strengthened significantly by the dilution of the line source 
function due to photon escape (such as the case of a simple two-level 
atom), while the increase of line-opacity due to overpopulation in 
the lower level of this transition is essential for the latter case.

\subsection{Comparison with Literature Data}

We now discuss the carbon abundances we have derived in 
comparison with previous studies taken from the literature. 

According to figure 4a, our [C/Fe] vs. [Fe/H] relation constructed
from 47 objects is characterized by 
(i) a gradual increase of [C/Fe] at a slightly supersolar level
with a decrease of [Fe/H] from $\sim 0$ to $\sim -1$, 
(ii) a subsolar [C/Fe] around $\sim -0.5$ (mostly giants) at 
$-3 \ltsim$~[Fe/H]~$\ltsim -2$, and 
(iii) upturn of [C/Fe] toward appearance of supersolar [C/Fe]
at the extremely metal-poor regime ($-4 \ltsim$~[Fe/H]~$\ltsim -3$).
Actually, these features are in good agreement with what has been
established so far; e.g., figure 6 of Norris, Ryan, and  Beers (1997) 
is quite similar to our figure 4a. Accordingly, we may state
that our analysis using C~{\sc i} 10683/10685/10691 lines has
yielded [C/Fe] results consistent with other previous work 
(where CH lines were mainly used in the very metal-poor regime) 
at least in the qualitative sense.

We then turn to the comparison of our [C/H] 
($\equiv A_{\rm C}^{\rm N} - A_{\rm C \odot}^{\rm N}$)
for individual stars with the literature data summarized in
table 3, which we collected mainly by invoking the SAGA database
(cf. section 1) and by adding data from several papers by ourselves 
(see the caption of table 3). We could thus get the published data 
for 37 stars ($\sim 80$\% of the total objects). 
These literature [C/H] values
are plotted by crosses (red $\cdots$ molecular lines mostly of CH, 
blue $\cdots$ C~{\sc i} lines) in figure 6, where our results 
are designated by circles
(open $\cdots$ giants, filled $\cdots$ dwarfs). Several notable 
characteristics can be read from this figure:\\
--- For disk stars with metallicity of $-1 \ltsim$~[Fe/H]
where C~{\sc i} lines are mostly used, the agreement is generally 
good, without any systematic discrepancy.\\
--- In contrast, at the very low metallicity range 
($-2.7 \ltsim$~[Fe/H]~$\ltsim -1$), we note that literature [C/H] values 
derived from CH lines (red crosses) tend to be appreciably lower 
(typically by several tenths dex) than our results, especially for 
low-gravity giants (open circles).\footnote{Actually, such a
CH vs. C~{\sc i} discrepancy is apparently observed in the results
of Tomkin et al. (1992), which is a rare study because two kinds of 
C abundances from C~{\sc i} lines and CH molecular band are given 
for a star. According to their table 4, the average difference
for 28 stars (in which both abundances are available) is
$\langle A({\rm CH}) - A({\rm C~I}) \rangle = -0.41$~dex,
which is nearly on the same order as we see in figure 6.
We should note, however, that the real discordance could be 
much more serious (even amounting up to $\gtsim 1$~dex), since 
$A$(CH) may further be significantly lowered (e.g., by a few tenths dex
to $\ltsim 1$~dex; especially at the very metal-poor condition) when 
new 3D model atmospheres are used (e.g., Asplund 2005; Collet 2007) 
because of the reduced temperature at the upper layers.}
Meanwhile, those derived from
C~{\sc i} lines (mostly 9061--9111 lines of multiplet 3) expressed 
by blue crosses do not show such a systematic tendency.\\
--- However, if we confine to extremely metal-poor stars at 
[Fe/H]~$\ltsim -3$, such a trend can not be clearly seen; e.g.,
CH-based literature [C/H]'s are even higher than our results
in HD~126587 and BD~+44~493. \\
--- Roughly summarizing, we may state that our C-abundances determined 
from C~{\sc i} 10683--10691 lines are almost consistent with the 
literature results if they are based on C~{\sc i} lines, while 
CH-based abundances published so far tend to be lower than ours
(especially for giants). 

%Figure 6
\setcounter{figure}{5}
\begin{figure}
  \begin{center}
    \FigureFile(70mm,100mm){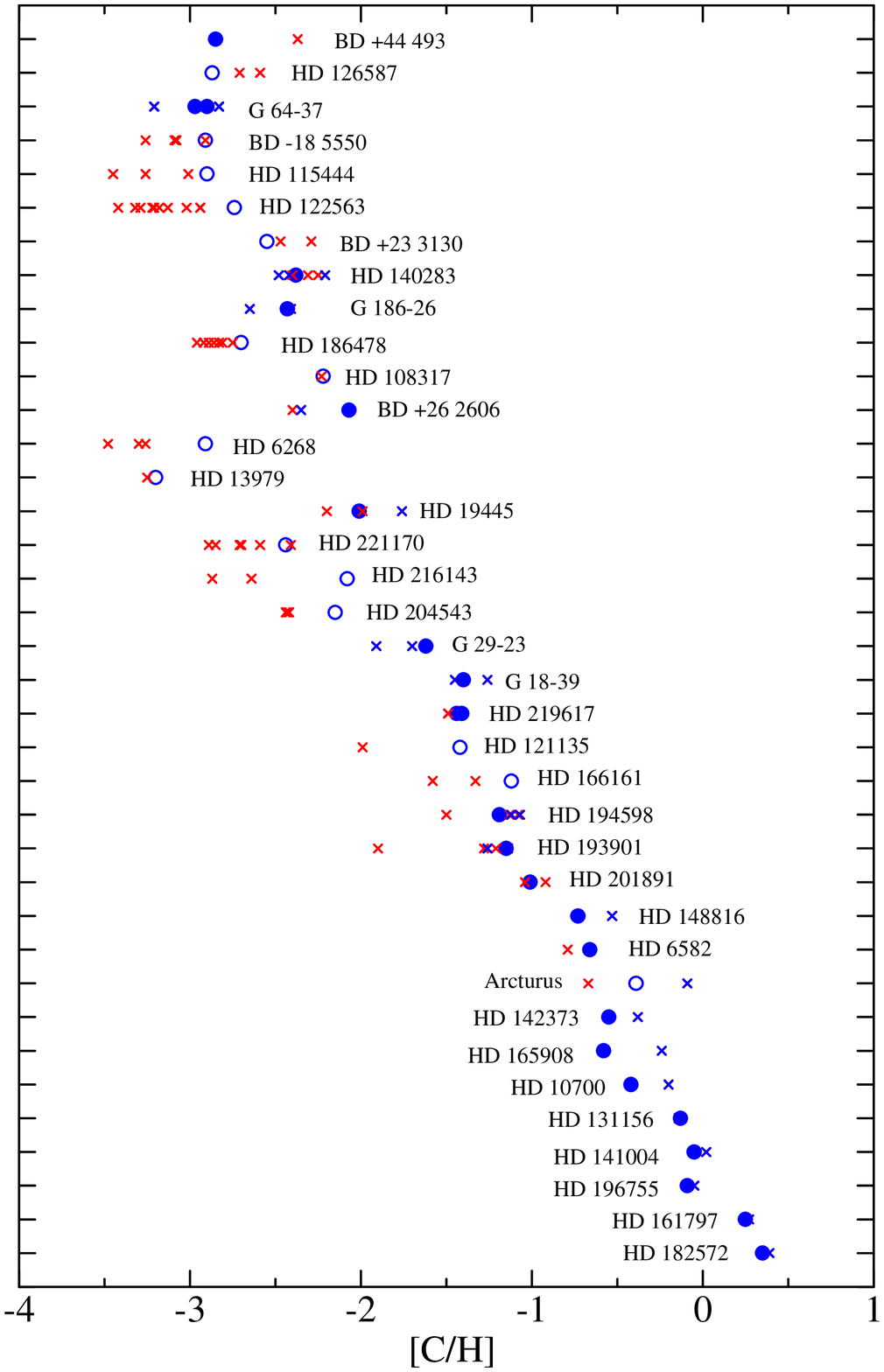}
    %%% \FigureFile(width,height){filename}
  \end{center}
\caption{
Comparison of [C/H] ($\equiv A_{\rm C} - A_{\rm C \odot}$)
values derived in this study (circles, where open and 
filled symbols denote giants and dwarfs, respectively) with those
taken from various published studies (crosses, where those derived from
molecular and atomic lines are discriminated in red and blue colors,
respectively), based on the data summarized in table 3. Stars are
arranged in the ascending order of [Fe/H] from top to bottom. 
}
\end{figure}

\subsection{Merits and Problems of C~I 1.07$\mu$m Lines}

Returning to the motivation of this study described in section 1, 
we again question to ourselves: Are C~{\sc i} 1.07~$\mu$m lines of 
multiplet 1 practically useful for stellar carbon abundance determinations, 
especially for very metal-poor stars? Our answer is as follows:
``Yes, definitely. We believe that these near-IR lines would 
promisingly be applied to investigate C abundances for a variety 
of stars. However, several things still remain to be done before
we can assure a sufficient precision in abundance determinations.''

We summarize below the favorable characteristic of these C~{\sc i} 
lines in comparison to other indicators.\\
--- These C~{\sc i} lines have sufficiently large strengths 
and would be visible down to extremely low-metallicity region. 
According to figure 5, if we could measure a very weak line 
of $EW \ltsim 10$~m$\rm\AA$ based on a high dispersion 
(e.g., $R\sim$~30000--50000) and high S/N (e.g., $\sim$~300--500) 
spectrum, $A_{\rm C}$ of a [Fe/H] $\sim -4$ star would be determinable
even for the case of [C/Fe]~$\sim 0$ (of course, detection becomes
much easier for carbon-enhanced stars of [C/Fe]~$>0$).\\ 
--- This merit almost equally applies to stars of any type, 
since whether a star is a giant or a dwarf does not significantly 
affect the line strength.
That is, since dwarfs with higher $\log g$ tend to 
have higher $T_{\rm eff}$ (e.g., turn-off stars with 
$T_{\rm eff} \sim 6000$~K and $\log g \sim 4$) and
giants with lower $\log g$ tend to have lower $T_{\rm eff}$ 
(e.g., evolved red giants with $T_{\rm eff} \sim 4000$~K and 
$\log g \sim 1$) as shown in figure 1a of Paper I,
these two opposite effect of $T_{\rm eff}$ and $\log g$ tend to
cancel with each other (cf. figure 5), which makes the line strength 
rather insensitive to $T_{\rm eff}$ as well as $\log g$.\\
--- Carbon abundances from CH molecular lines are very sensitive 
to $T_{\rm eff}$ (cf. table 5 of Tomkin et al. 1992), as well as 
to the temperature structure in the upper layer and to the 3D
effect (e.g., Asplund 2005; Behara et al. 2008). In this respect,
C~{\sc i} lines are much less sensitive to such effects.\\
--- These C~{\sc i} 10683--10691 lines of multiplet 1 are 
more advantageous than the C~{\sc i} 9061--9111 lines of 
multiplet 3, because they are stronger and situate in a 
spectral region almost free from telluric lines.

Despite these merits, however, we have to keep in mind that 
problems still remain in deriving the C abundances from these
C~{\sc i} lines, especially when a sufficient precision is pursued.
Above all, given that these lines are subject to rather a large
non-LTE effect, attention should be paid to use as reliable
non-LTE abundance corrections as possible. Here, the difficult
problem is the treatment of collisions with neutral hydrogen atoms,
or how to choose the value of $k$ in equation (1). Since the non-LTE 
corrections appreciably depend upon this parameter (as we can see
from figure 4c), its adequate choice is mandatory. 

Yet, we consider that $k=1$ adopted in this investigation (as in 
our past studies) is reasonable for the following reasons:\\ 
--- If we use $k=0.1$, for example, we would obtain   
$A_{\rm C,\odot}^{\rm N} = 8.40$ ($\Delta = -0.28$) as the solar
carbon abundance (instead of 8.54 for $k=1$). Although this value 
is almost the same as the most recent solar C abundance of 8.43 
claimed by Asplund et al. (2009) using the state-of-the-art 3D model, 
our result based on the classical 1D model should not be directly 
compared to it. Considering that Takeda and Honda (2005) 
derived the solar C abundances of  8.49, 8.76, and 8.70 
(based on the same solar model as used in this study) from 
three non-LTE insensitive lines ([C~{\sc i}]~8727, C~{\sc i}~5052, 
and C~{\sc i}~5380), we feel reluctant to further lower 
the value of 8.54.\\
--- Admittedly, the problematic discrepancy between our [C/H] 
results and those from CH molecular lines seen in metal-poor giants 
(cf. subsection 4.2) may significantly be mitigated by reducing $k$. 
If we do so, however, our C~{\sc i}-based abundances for dwarfs
would be overcorrected (actually, non-LTE corrections for dwarfs are
more sensitive to $k$ than those of giants; cf. figure 4d) to become 
discordant with the literature results, since the consistency between
CH and C~{\sc i} results are not so bad in metal-poor dwarfs 
unlike the case for giants (cf. figure 6). 
We rather suspect that CH-based abundances of giants may 
have been incorrect. Given that modeling of giants atmosphere 
is generally less reliable than the case of dwarfs, previously 
published C abundances of metal-poor giants based on CH lines 
may have been appreciably underestimated due to some imperfectness 
of the atmospheric model, reflecting the enormous sensitivity 
of molecular lines to errors in the temperature structure.
For example, since the ubiquitous existence of chromospheric
activities even in very metal-poor old stars is evidenced
by the detection of He~{\sc i}~10830 line (cf. Takeda \& Takada-Hidai 
2011b), the temperature of molecule-forming upper layers may be
higher than that predicted by classical model atmospheres
or UV radiation coming from the chromosphere may promote non-thermal
dissociation of molecules in the photosphere, which would lead 
to underestimation of carbon abundances derived from CH 
molecular band (if based on the conventional method of analysis).

In any event, it is very important to determine the value of
$k$ in an empirical manner. One way for doing this may be to 
require the abundance consistency between different lines, 
such as done by Takeda and Honda (2005, cf. appendix~1 therein)
for oxygen lines.
Actually we challenged it by comparing the abundances from lines of 
multiplet 1 and multiplet 3, as described in the Appendix. 
Unfortunately, this trial turned out unsuccessful, because
the $k$-sensitivity of these two line groups were essentially
the same. As an alternative approach, it would be promising to 
compare the center-to-limb variation of C~{\sc i} 10683--10691 lines
on the solar disk with the simulations done for various $k$ values, 
as recently tried by Pereira, Kiselman, and Asplund (2009a) 
as well as Pereira, Asplund, and Kiselman (2009b) for the oxygen lines.
We would be able to use these near-IR C~{\sc i} lines for precise
C abundance determinations with confidence, when the calibration 
of this collisional parameter has been established.

%Sect. 5
\section{Conclusion}

Motivated by the fact that C~{\sc i} lines at $\sim$~1.07~$\mu$m
(multiplet 1) have barely been exploited for stellar C abundance
studies so far, despite their sufficient strengths and being in a
favorable spectral region free from telluric lines, we 
carried out a non-LTE spectrum-fitting analysis of 
C~{\sc i} 10683--19691 lines for selected 46 halo/disk stars 
in a wide metallicity range ($\sim -3.7$~[Fe/H]~$\ltsim +0.3$) 
based on the $zJ$-band spectral data collected with IRCS+AO188 of 
the Subaru Telescope, with an aim of examining whether the 
C abundances can be successfully determined by these lines
especially for very metal-poor stars where lines of CH
molecules have been mainly invoked.
 
These C~{\sc i} lines were confirmed to be clearly visible,
from which C abundances could be successfully determined for 
all stars down to [Fe/H]~$\sim -3.7$. The resulting 
[C/Fe] vs. [Fe/H] diagram revealed the well-known trend
established by previous studies; such as a gradual increase of 
[C/Fe] with a decrease of [Fe/H] from $\sim 0$ to $\sim -1$, 
subsolar [C/Fe] of $\sim -0.5$ at $-3 \ltsim$~[Fe/H]~$\ltsim -2$,
and upturn of [C/Fe] toward appearance of supersolar [C/Fe]
at $-4 \ltsim$~[Fe/H]~$\ltsim -3$. We may thus regard that
our analysis is consistent with previous studies at least 
qualitatively.

We also compared the resulting [C/H] for individual stars with the 
published data collected from various literature (data for 37 stars
were available). While the agreement is mostly good for disk stars 
($-1 \ltsim$~[Fe/H]) where C~{\sc i} lines are generally used, 
we found that the literature [C/H] values of very metal-poor stars
($-2.7 \ltsim$~[Fe/H]~$\ltsim -1$) derived from CH lines 
tend to be appreciably lower than our results especially for 
low-gravity giants, though those derived from C~{\sc i} 9061--9111 
lines of multiplet 3 do not show such a systematic tendency.
The reason for this disagreement is yet to be investigated.

Since the abundances of these neutral C lines are subject to 
an appreciably large non-LTE effect (typically by several tenths dex) 
whose importance progressively grows as the metallicity is lowered, 
attention should be paid to the reliability of non-LTE corrections
to be applied. Here, the difficult problem is how to treat 
collisions with neutral hydrogen atoms, for which classical rates 
are often multiplied by a reduction factor $k (\le 1)$. 
Although we consider that the choice of $k=1$ (standard classical 
treatment) is still reasonable, its choice has a significant 
influence on the extent of non-LTE effect (e.g., if we use $k=0.1$, 
non-LTE corrections would become twice as large as the $k=1$ case).
Thus, establishing $k$ in some empirical manner (e.g., by using the
center-to-disk variation on the solar disk) is urgently awaited.

\bigskip

One of the authors (M. T.-H.) is grateful for a financial support 
from a grant-in-aid for scientific research (C, No. 22540255) 
from the Japan Society for the Promotion of Science.

This research has made extensive use of the SAGA database system, 
maintained by Takuma Suda, Yutaka Katsuta, and Shimako Yamada. 

%Appendix (figure 6, figure 7)
\appendix
\section*{Non-LTE Corrections of C~I 9061--9111 Lines Revisited}

Takeda and Honda (2005; cf. appendix 2 therein) computed an extensive 
grid of non-LTE equivalent widths and abundance corrections for five 
C~{\sc i} lines of multiplet 3 (3s~$^{3}{\rm P}^{\rm o}$ -- 3p~$^{3}{\rm P}$) 
at 9061.44, 9078.29, 9088.51, 9094.83, and 9111.81~$\rm\AA$. They also 
carried out a non-LTE reanalysis of published equivalent-width data
of these C~{\sc i}~9061--9111 lines along with those of 
C~{\sc i}~7111--7119 lines (multiplet 26, 25.02). 

Having reexamined those previous calculations, 
we realized that there was a mis-assignment of departure 
coefficients in computing the non-LTE equivalent widths 
of C~{\sc i}~9061--9111 lines. More precisely, we had erroneously
used the departure coefficients corresponding to multiplet 1
(3s~$^{3}{\rm P}^{\rm o}$ -- 3p~$^{3}{\rm D}$) for
these multiplet 3 lines. Accordingly, we decided to redo the 
calculations (this time, computations were performed not only for 
$k=1$ as assumed before, but also for $k=0.1$), and the resulting 
grids of new non-LTE corrections ($\Delta$) and equivalent-widths 
($EW$) are presented in electronic table E1 along with 
those of C~{\sc i}~10685--10729 lines.

Comparing the new results with the old ones, we realized that
the differences are insignificant; actually, the change of $\Delta$
is $\ltsim 0.1$~dex at most, as shown in figure 7.
This is due to the fortunate fact that multiplets 1 and 3
have the same lower term (3s~$^{3}{\rm P}^{\rm o}$),
while the upper terms (3p~$^{3}{\rm D}$ and 3p~$^{3}{\rm P}$) 
have similar departure coefficients because they lie closely 
to each other. Accordingly, we do not see any need to apply
significant changes to what is described in appendix 2 of
Takeda and Honda (2005), which still remain essentially valid
in the practical sense.

%Figure 7
\setcounter{figure}{6}
\begin{figure}
  \begin{center}
    \FigureFile(70mm,100mm){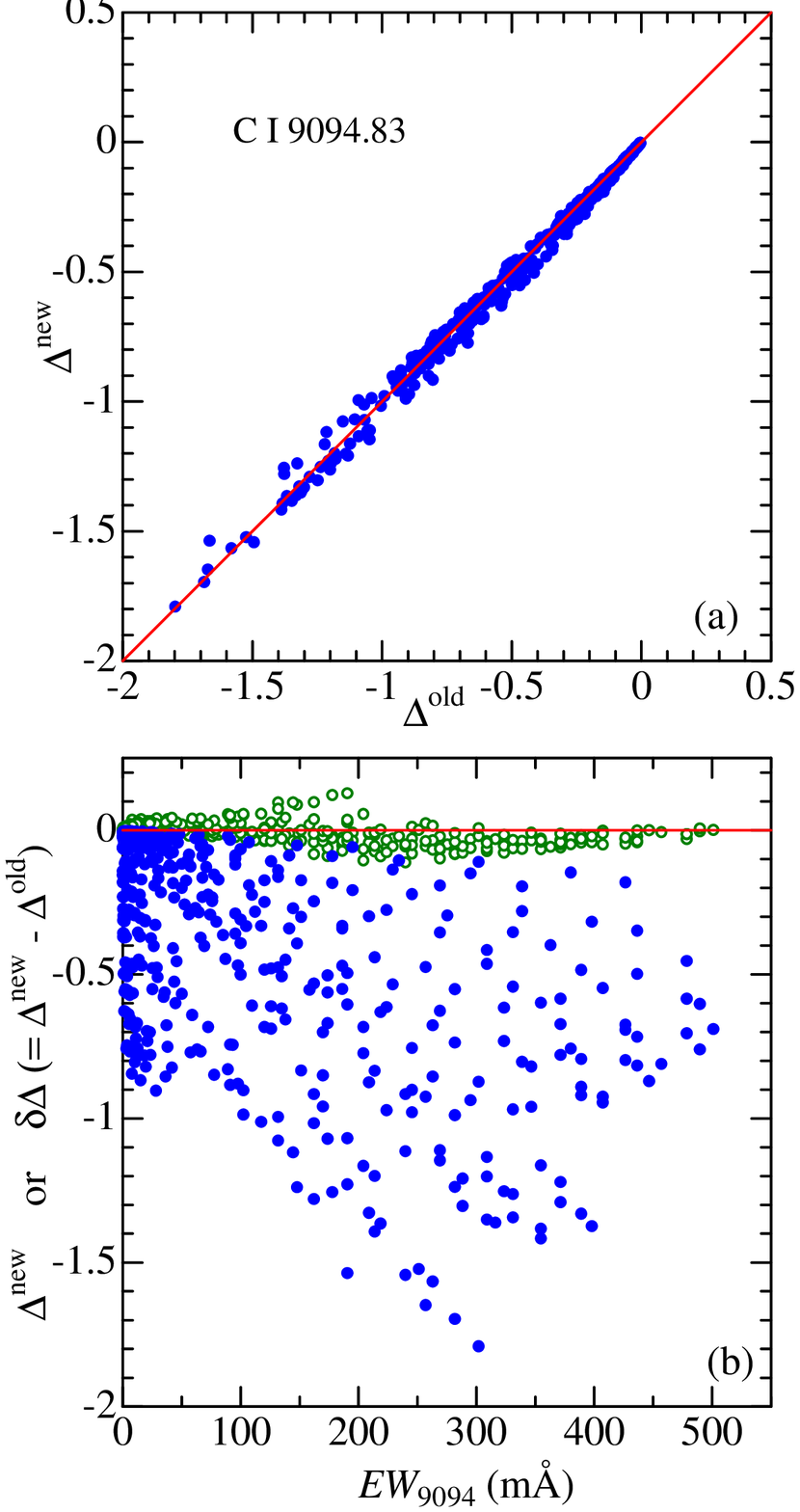}
    %%% \FigureFile(width,height){filename}
  \end{center}
\caption{
(a) Comparison of the old non-LTE corrections ($\Delta^{\rm old}$) for 
the 9094.83 line (the strongest one of the five 9061--9111 lines of 
RMT~3) computed by Takeda and Honda (2005; cf. appendix 2 therein) 
for an extensive grid of model atmospheres, for which we found 
that departure coefficients for the upper term had been 
erroneously assigned by mistake, with the new results ($\Delta^{\rm new}$) 
correctly recomputed in this study. Note that these calculations 
correspond to $k=1$ (standard treatment of neutral hydrogen collisions). 
(b) Similarly to panel (a), $\Delta^{\rm new}$ (filled symbols) and 
the difference $\delta \Delta$ ($\equiv \Delta^{\rm new} - \Delta^{\rm old}$; 
open symbols) are plotted against the (non-LTE) equivalent width 
($EW_{9094}$).
}
\end{figure}

%Figure 8
\begin{figure}
  \begin{center}
    \FigureFile(70mm,100mm){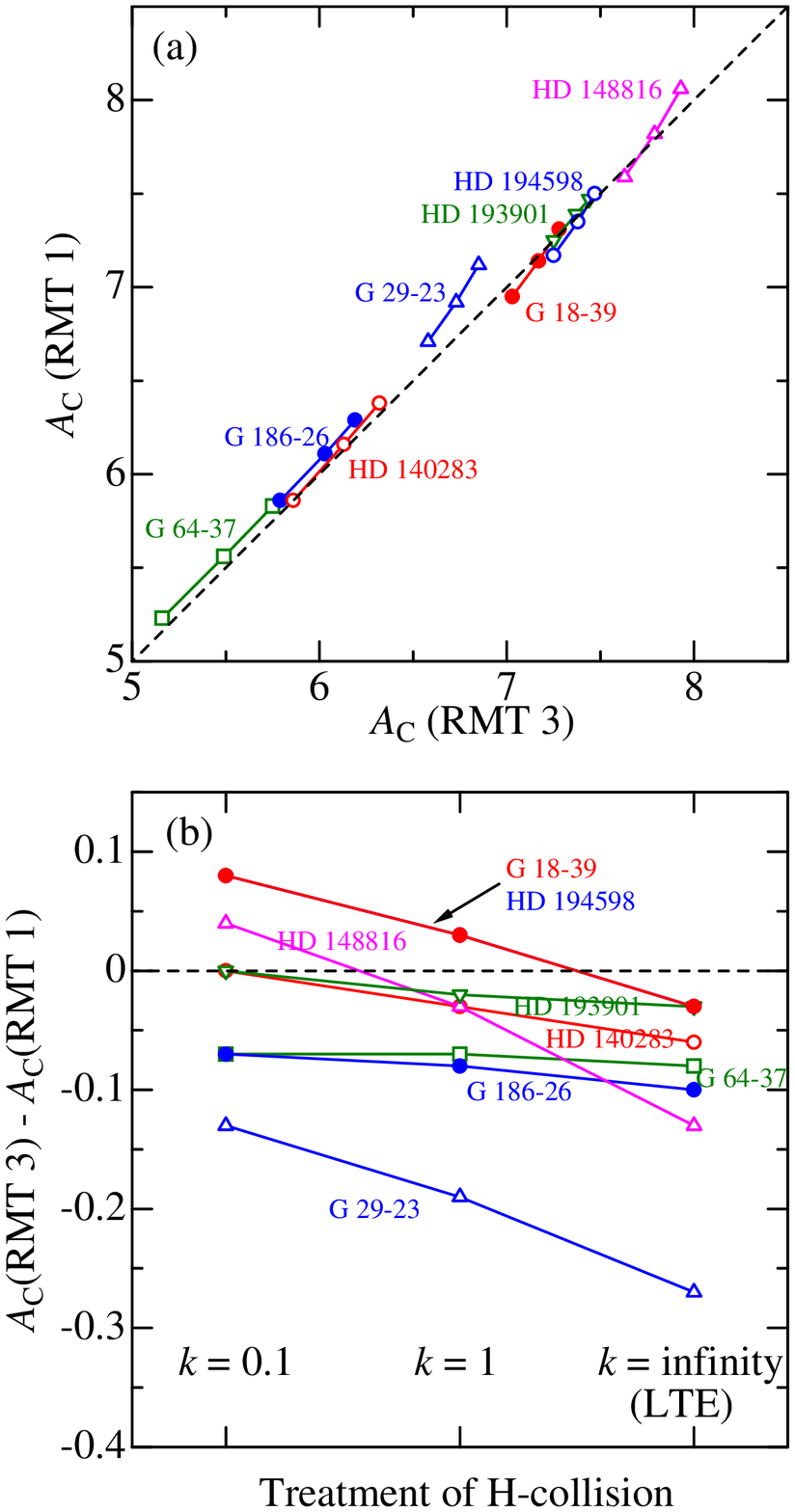}
    %%% \FigureFile(width,height){filename}
  \end{center}
\caption{
(a) Comparison of the carbon abundances ($A_{\rm C}$) of 8 stars derived from 
this study (C~{\sc i}~10683--10691 lines of RMT~1) with those obtained by
reanalyzing Akerman et al.'s (2004) $EW$ data of C~{\sc i}~9061--9111 
lines (RMT~3), where three different conditions were adopted
regarding how neutral hydrogen collisions ($C_{\rm H}$) are taken into account: 
(i) LTE ($k = \infty$), (ii) standard treatment ($k=1$), and (iii) reduction 
of $C_{\rm H}$ by 1/10 ($k=0.1$); note that 
$A_{\rm C}({\rm i})>A_{\rm C}({\rm ii})>A_{\rm C}({\rm iii})$ generally holds.
(b) Graphical display of how $A_{\rm C}$(RMT 3)~$-$~$A_{\rm C}$(RMT 1)
varies with a different choice in the treatment of $C_{\rm H}$.
}
\end{figure}

In connection with the newly recomputed non-LTE corrections
for multiplet 3 lines, we tried to empirically determine the 
value of $k$ defined in equation (1) by requiring the consistency 
between the carbon abundances derived from C~{\sc i}~10685--10691 
lines (obtained from this study) and those from 
C~{\sc i} 9061--9111 lines (based on reanalysis of published
equivalent widths). By consulting the work of Akerman et al. (2004),  
$EW$s of 9061--9111 lines for 8 stars were available, from which
we derived the mean abundance of each star for three choices of $k$:
$k =0.1$, $k=1$, and $k=\infty$ (i.e., LTE). Comparison of 
such computed $A_{\rm C}$(9061--9111:RMT~3) with our
$A_{\rm C}$(10685--10691:RMT~1) is presented in figure 8a.
Unfortunately, we can recognize from this figure that the 
$k$-sensitivity of both multiplet lines is nearly the same
with each other, which means that empirically establishing $k$
by the requirement of $A_{\rm C}$(RMT~1) = $A_{\rm C}$(RMT~3) is
hardly practicable (cf. figure 8b). We would have to find
some other different carbon lines to be compared, in order to 
accomplish this aim.

%\clearpage

\onecolumn 

%Figure 2
\setcounter{figure}{1}
\begin{figure}
  \begin{center}
    \FigureFile(160mm,230mm){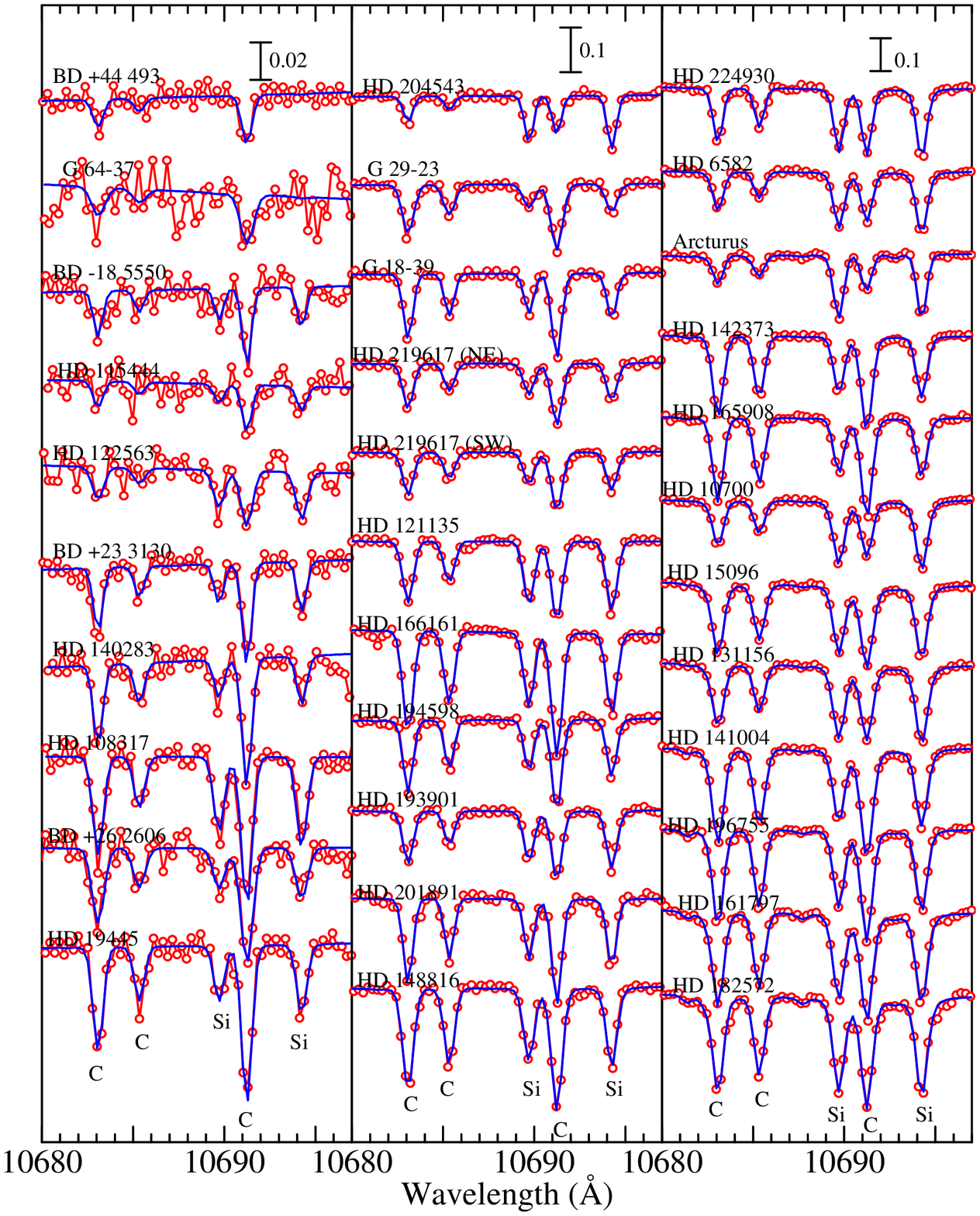}
    %%% \FigureFile(width,height){filename}
  \end{center}
\caption{
Synthetic spectrum fitting in the 10680--10697~$\rm\AA$ region, 
comprising three C~{\sc i} and two Si~{\sc i} lines, for 33 stars in 
the wide range of metallicity based on the 2009 July data (cf. Paper I). 
The best-fit theoretical spectra are shown by (blue) 
solid lines, while the observed data are plotted by (red) circles.  
In each panel (from left to right), the spectra are arranged 
(from top to bottom) in the ascending order of [Fe/H] 
as in table 1. An appropriate vertical offset (0.05, 0.2, 0.25 for 
the left, middle, and right panel, respectively) is applied 
to each spectrum relative to the adjacent one. 
The wavelength scale of each spectrum is adjusted to the
laboratory frame. }
\end{figure}

%Figure 3
\setcounter{figure}{2}
\begin{figure}
  \begin{center}
    \FigureFile(160mm,160mm){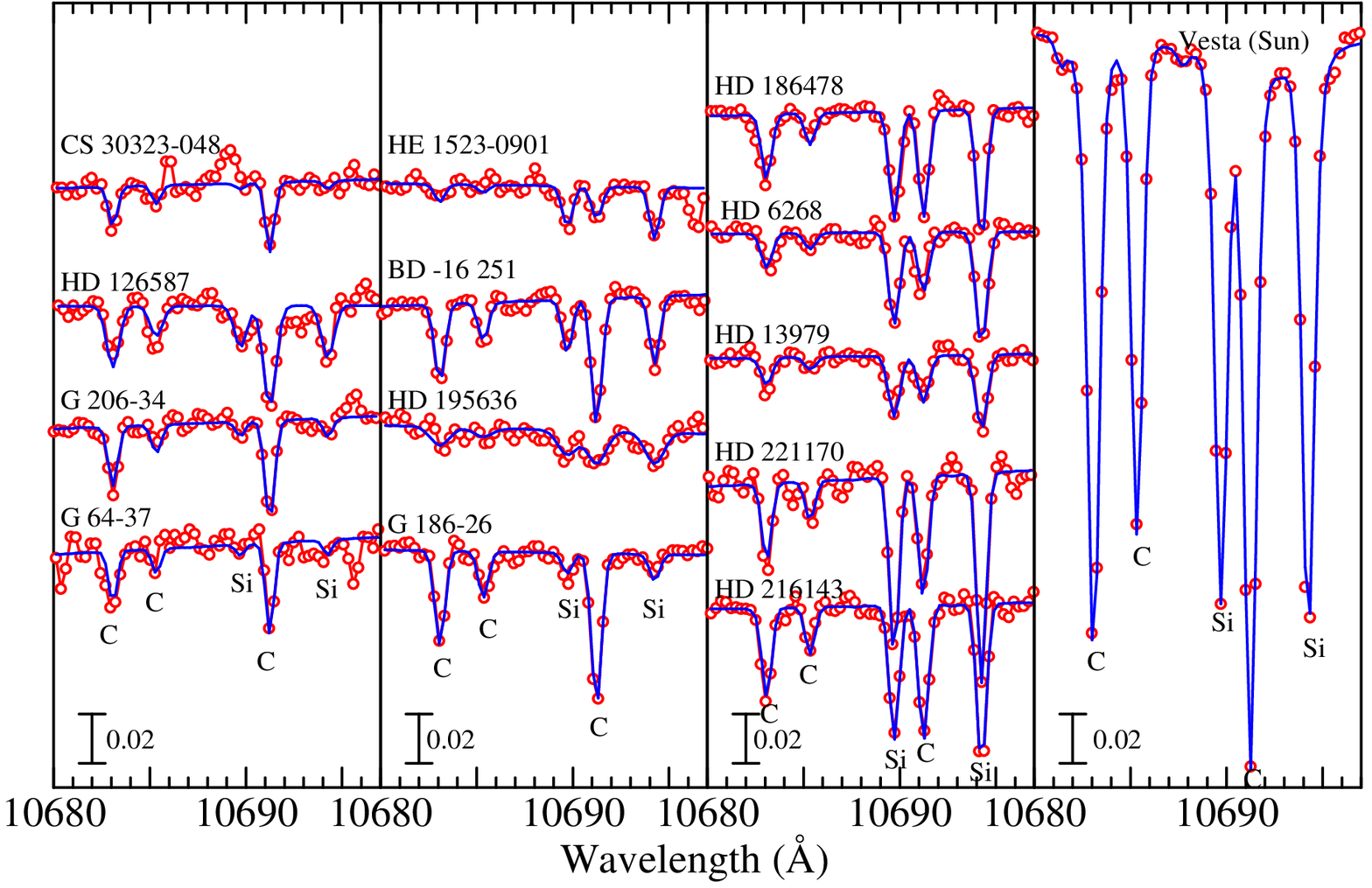}
    %%% \FigureFile(width,height){filename}
  \end{center}
\caption{
Synthetic spectrum fitting in the 10680--10697~$\rm\AA$ region for 
12 very metal-poor stars (along with Vesta) based on the 2011 August 
data (cf. Paper II). The same value (0.05) of vertical offset is 
applied to each spectrum relative to the adjacent one. 
Otherwise, the same as in figure 2.}
\end{figure}

%Table 1
\clearpage
\setcounter{table}{0}
\begin{table}[h]
\caption{Parameters of the program stars and the results of the abundance analysis.}
\scriptsize
\begin{center}
\begin{tabular}{cc@{ }c@{ }c@{ }c crccl} 
\hline\hline
Name & $T_{\rm eff}$ & $\log g$ & $v_{\rm t}$ & [Fe/H] & $A_{\rm C}^{\rm N}$ & 
$EW_{10691}$ & $\Delta_{10691}$ & [C/Fe] & Remark \\
 & (K) & (cm~s$^{-2}$) & (km~s$^{-1}$) & (dex) & (dex) & (m$\rm\AA$) &(dex)&(dex)& \\
\hline
\multicolumn{10}{c}{(2009 July data)}\\
BD~+44~493   &5510 & 3.70& 1.30& $-$3.68&  5.69&  19.6& $-$0.38& +0.83& \\
G~64-37      &6432 & 4.24& 1.50& $-$3.08&  5.64&  25.7& $-$0.29& +0.18& also observed in 2011 August\\
BD~$-$18~5550  &4750 & 1.40& 1.80& $-$3.06&  5.63&  26.4& $-$0.34& +0.15& \\
HD~115444    &4721 & 1.74& 2.00& $-$2.71&  5.64&  16.9& $-$0.24& $-$0.19& \\
HD~122563    &4572 & 1.36& 2.90& $-$2.61&  5.80&  23.7& $-$0.24& $-$0.13& \\
BD~+23~3130  &5000 & 2.20& 1.40& $-$2.60&  5.99&  36.6& $-$0.28& +0.05& \\
HD~140283    &5830 & 3.67& 1.90& $-$2.55&  6.16&  47.0& $-$0.25& +0.17& \\
HD~108317    &5310 & 2.77& 1.90& $-$2.35&  6.32&  60.9& $-$0.29& +0.13& \\
BD~+26~2606  &5875 & 4.10& 0.40& $-$2.30&  6.47&  52.3& $-$0.16& +0.23& \\
HD~19445     &6130 & 4.39& 2.10& $-$2.05&  6.53&  62.3& $-$0.13& +0.04& \\
HD~204543    &4672 & 1.49& 2.00& $-$1.72&  6.39&  55.0& $-$0.25& $-$0.43& \\
G~29$-$23      &6194 & 4.04& 1.50& $-$1.69&  6.92& 119.8& $-$0.27& +0.07& \\
G~18$-$39      &6093 & 4.19& 1.50& $-$1.46&  7.14& 131.9& $-$0.23& +0.06& \\
HD~219617NE  &5825 & 4.30& 1.40& $-$1.40&  7.13& 104.5& $-$0.13& $-$0.01& North--East component\\
HD~219617SW  &5825 & 4.30& 1.40& $-$1.40&  7.10& 101.2& $-$0.13& $-$0.04& South--West component\\
HD~121135    &4934 & 1.91& 1.60& $-$1.37&  7.12& 133.3& $-$0.44& $-$0.05& \\
HD~166161    &5350 & 2.56& 2.25& $-$1.22&  7.42& 206.3& $-$0.57& +0.10& \\
HD~194598    &6020 & 4.30& 1.40& $-$1.15&  7.35& 146.0& $-$0.20& $-$0.04& \\
HD~193901    &5699 & 4.42& 1.20& $-$1.10&  7.39& 118.1& $-$0.11& $-$0.05& \\
HD~201891    &5900 & 4.19& 1.40& $-$1.10&  7.53& 167.1& $-$0.23& +0.09& \\
HD~148816    &5860 & 4.07& 1.60& $-$1.00&  7.81& 215.3& $-$0.29& +0.27& \\
HD~224930    &5275 & 4.10& 1.05& $-$0.90&  8.02& 153.7& $-$0.11& +0.38& \\
HD~6582      &5331 & 4.54& 0.73& $-$0.81&  7.88& 122.6& $-$0.06& +0.15& \\
Arcturus    &4281 & 1.72& 1.49& $-$0.55&  8.15&  73.0& $-$0.23& +0.16& \\
HD~142373    &5776 & 3.83& 1.26& $-$0.51&  7.99& 228.3& $-$0.30& $-$0.04& \\
HD~165908    &6183 & 4.35& 1.24& $-$0.46&  7.96& 243.3& $-$0.25& $-$0.12& \\
HD~10700     &5420 & 4.68& 0.66& $-$0.43&  8.12& 148.3& $-$0.06& +0.01& \\
HD~15096     &5375 & 4.30& 0.80& $-$0.20&  8.70& 246.0& $-$0.11& +0.36& \\
HD~131156    &5527 & 4.60& 1.10& $-$0.13&  8.41& 208.0& $-$0.09& +0.00& \\
HD~141004    &5877 & 4.11& 1.17& $-$0.01&  8.49& 305.0& $-$0.22& $-$0.04& \\
HD~196755    &5750 & 3.83& 1.23&  0.09&  8.45& 282.5& $-$0.26& $-$0.18& \\
HD~161797    &5580 & 3.99& 1.11&  0.29&  8.79& 302.7& $-$0.19& $-$0.04& \\
HD~182572    &5566 & 4.11& 1.07&  0.33&  8.89& 315.2& $-$0.17& +0.02& \\
\hline
\multicolumn{10}{c}{(2011 August data)}\\
CS~30323-048 &6338 & 4.32& 1.50& $-$3.21&  5.52&  17.6& $-$0.28& +0.19& \\
HD~126587    &4700 & 1.05& 1.70& $-$3.16&  5.67&  34.5& $-$0.39& +0.29& \\
G~206-34     &5825 & 3.99& 1.50& $-$3.12&  5.93&  27.3& $-$0.25& +0.51& \\
G~64-37      &6432 & 4.24& 1.50& $-$3.08&  5.57&  22.0& $-$0.28& +0.11& also observed in 2009 July\\
HE~1523$-$0901 &4630 & 1.00& 2.60& $-$2.95&  5.07&   8.9& $-$0.28& $-$0.52& \\
BD~$-$16~251   &4825 & 1.50& 1.80& $-$2.91&  5.78&  35.6& $-$0.35& +0.15& \\
HD~195636    &5370 & 2.40& 1.50& $-$2.77&  5.30&  16.8& $-$0.33& $-$0.47& \\
G~186-26     &6417 & 4.42& 1.50& $-$2.54&  6.11&  44.3& $-$0.19& +0.11& \\
HD~186478    &4730 & 1.50& 1.80& $-$2.42&  5.84&  29.7& $-$0.27& $-$0.28& \\
HD~6268      &4735 & 1.61& 2.10& $-$2.30&  5.63&  17.8& $-$0.21& $-$0.61& \\
HD~13979     &5075 & 1.90& 1.30& $-$2.26&  5.34&  14.8& $-$0.23& $-$0.94& \\
HD~221170    &4560 & 1.37& 1.60& $-$2.00&  6.10&  32.4& $-$0.21& $-$0.44& \\
HD~216143    &4525 & 1.77& 1.90& $-$1.92&  6.46&  37.0& $-$0.17& $-$0.16& \\
Vesta        &5780 & 4.44& 1.00&  0.00&  8.54& 283.7& $-$0.14& +0.00& substitute for the Sun\\
\hline
\end{tabular}
\end{center}
In columns 1 through 5 are given the star designation, 
effective temperature, logarithmic surface gravity, 
microturbulent velocity dispersion, and Fe abundance 
relative to the Sun, which are the same as adopted in Papers I and II.
Columns 6--9 present the results of the abundance analysis.
$A_{\rm C}^{\rm N}$ is the non-LTE logarithmic abundance of C 
(in the usual normalization of $A_{\rm H}$ = 12.00) derived from 
spectrum-synthesis fitting, $EW_{10691}$ is the equivalent width 
(in m$\rm\AA$) for the C~{\sc i} 10691 line inversely computed 
from $A_{\rm C}^{\rm N}$, $\Delta_{10691}$ is the non-LTE correction 
($\equiv A_{\rm C}^{\rm N} - A^{\rm L}_{10691}$) for the 
C~{\sc i} 10691 line, and [C/Fe] ($\equiv A_{\rm C}^{\rm N} - 8.54 - $~[Fe/H]) 
is the C-to-Fe logarithmic abundance ratio relative to the Sun. 
In each of the two groups corresponding to different observational epochs,
the data are arranged in the ascending order of [Fe/H]. 
\end{table}

%Table 2
\clearpage
\setcounter{table}{1}
\begin{table}[h]
\caption{Atomic data of important lines relevant for this study.}
\scriptsize
\begin{center}
\begin{tabular}
{cccrcrccc}\hline \hline
Species & RMT & Transition & $\lambda$ ($\rm\AA)$ & $\chi_{\rm low}$ (eV) & $\log gf$ & Gammar & Gammas & Gammaw \\
 \hline
  C~{\sc i}& 3 & 3s~$^{3}{\rm P}^{\rm o}_{3}$ -- 3p $^{3}{\rm P}_{5}$ & 9061.436 &  7.483 &  $-0.335$ &   (7.43) &  $-5.32$ &  ($-7.59$) \\  
  C~{\sc i}& 3 & 3s~$^{3}{\rm P}^{\rm o}_{3}$ -- 3p $^{3}{\rm P}_{3}$ & 9078.288 &  7.483 &  $-0.557$ &   (7.43) &  $-5.32$ &  ($-7.59$) \\  
  C~{\sc i}& 3 & 3s~$^{3}{\rm P}^{\rm o}_{3}$ -- 3p $^{3}{\rm P}_{1}$ & 9088.513 &  7.483 &  $-0.432$ &   (7.43) &  $-5.32$ &  ($-7.59$) \\  
  C~{\sc i}& 3 & 3s~$^{3}{\rm P}^{\rm o}_{5}$ -- 3p $^{3}{\rm P}_{5}$ & 9094.830 &  7.488 &   +0.142 &   (7.43) &  $-5.32$ &  ($-7.59$) \\  
  C~{\sc i}& 3 & 3s~$^{3}{\rm P}^{\rm o}_{5}$ -- 3p $^{3}{\rm P}_{3}$ & 9111.807 &  7.488 &  $-0.335$ &   (7.43) &  $-5.32$ &  ($-7.59$) \\  
  C~{\sc i}& 1 & 3s~$^{3}{\rm P}^{\rm o}_{3}$ -- 3p $^{3}{\rm D}_{5}$ & 10683.083 &  7.483 &  +0.076 &   (7.29) &  $-5.40$ &  ($-7.62$) \\  
  C~{\sc i}& 1 & 3s~$^{3}{\rm P}^{\rm o}_{1}$ -- 3p $^{3}{\rm D}_{3}$ & 10685.343 &  7.480 &  $-0.276$ &   (7.29) &  $-5.40$ &  ($-7.62$) \\  
  C~{\sc i}& 1 & 3s~$^{3}{\rm P}^{\rm o}_{5}$ -- 3p $^{3}{\rm D}_{7}$ & 10691.240 &  7.488 &   +0.348 &   (7.29) &  $-5.40$ &  ($-7.61$) \\  
  C~{\sc i}& 1 & 3s~$^{3}{\rm P}^{\rm o}_{3}$ -- 3p $^{3}{\rm D}_{3}$ & 10707.317 &  7.483 &  $-0.401$ &   (7.29) &  $-5.40$ &  ($-7.62$) \\  
  C~{\sc i}& 1 & 3s~$^{3}{\rm P}^{\rm o}_{5}$ -- 3p $^{3}{\rm D}_{5}$ & 10729.530 &  7.488 &  $-0.401$ &   (7.29) &  $-5.40$ &  ($-7.62$) \\  
 Si~{\sc i}&53 & 4p~$^{3}{\rm D}_{3}$ -- 4d $^{3}{\rm F}^{\rm o}_{5}$ & 10689.716 &  5.954 &  $-0.190$ &   (7.29) &  $-4.23$ &  ($-7.29$) \\  
 Si~{\sc i}&53 & 4p~$^{3}{\rm D}_{3}$ -- 4d $^{5}{\rm F}^{\rm o}_{7}$ & 10694.252 &  5.964 &  $-0.060$ &   (7.29) &  $-4.23$ &  ($-7.29$) \\  
\hline
\end{tabular}
\end{center}
All data are were taken from Kurucz and Bell's (1995) compilation
as far as available. RMT is the multiplet number given by the 
Revised Multiplet Table (Moore 1959).
In the last three columns are given the damping parameters in the c.g.s. unit:\\
Gammar is the radiation damping constant (s$^{-1}$), $\log\gamma_{\rm rad}$.
Gammas is the Stark damping width per electron density (cm$^{-3}$)
at $10^{4}$ K, $\log(\gamma_{\rm e}/N_{\rm e})$.
Gammaw is the van der Waals damping width per hydrogen density (cm$^{-3}$)
at $10^{4}$ K, $\log(\gamma_{\rm w}/N_{\rm H})$. 
Note that the values in parentheses are the default damping parameters 
computed within the Kurucz's WIDTH program (cf. Leusin, Topil'skaya 1987),
because of being unavailable in Kurucz and Bell (1995).
The meanings of other columns are self-explanatory.
\end{table}

%Table 3
\setcounter{table}{2}
\scriptsize
\setlength{\tabcolsep}{3pt}
\begin{longtable}{crccrrcl}
\caption{Comparison of the [C/H] results and atmospheric parameters 
with the literature values.}
%\begin{center}
%\begin{tabular}{c rccrr cl}
\hline\hline
Star & $T_{\rm eff}$ & $\log g$ & $v_{\rm t}$ & [Fe/H] & [C/H] &Ref. & Line\\
  & (K) & (cm~s$^{-1}$) & (km~s$^{-1}$) & (dex) & (dex) &  & \\
\hline
\endhead
\hline
\endfoot
%%% beginning of foot in last page
\hline
\multicolumn{7}{l}{\hbox to 0pt{\parbox{150mm}{\footnotesize
For the metal-poor objects with [Fe/H]~$\ltsim -0.8$ (i.e., from BD~+44~493
through HD~6582), we consulted the SAGA database (Suda et al. 2008, 2011)
as of the end of 2012 November, in which the [C/H] values are reduced 
to the reference solar carbon abundance of 8.55 (Grevesse et al. 1996).
In addition, we also included [C/H] values from two papers which were 
missing in SAGA: 5 stars (HD~19445, HD~140283, HD~193901, HD~194598, 
BD~+26~2606) from Tomkin et al. (1992) and 5 stars (G~64-37, HD~140283, 
G~29-23, G~18-39, G~186-26) from Fabbian et al. (2009; results for
the $S_{\rm H}=1$ case were adopted and renormalized to the solar
abundance of 8.55). Meanwhile, regarding stars of 
[Fe/H]~$\gtsim -0.5$ (i.e., from Arcturus through HD~182572) where 
SAGA does not cover, the data were taken from Kjaergaard et al. (1982) 
and Ram\'{\i}rez et al. (2011) for Arcturus, as well as from Takeda 
and Honda (2005) for the other disk stars (an extensive literature 
survey has not been performed for those non-SAGA stars).
The reference source is given in the 7th column, and the adopted lines 
used for C-abundance determination are also noted in the 8th column.
Our data adopted/derived in this study are expressed in Italic.
Stars are arranged in the ascending order of [Fe/H] (the metallicity
value adopted in this study).\\ 
\newline
Key to the references: 
AKE04 --- Akerman et al. (2004); 
AOK02 --- Aoki et al. (2002); 
AOK05 --- Aoki et al. (2005); 
AOK07 --- Aoki et al. (2007); 
AOK08 --- Aoki and Honda (2008); 
BAR05 --- Barklem et al. (2005); 
CAY04 --- Cayrel et al. (2004); 
FAB09 --- Fabbian et al. (2009); 
GRA00 --- Gratton et al. (2000); 
HON04 --- Honda et al. (2004); 
ITO09 --- Ito et al. (2009); 
IVA06 --- Ivans et al. (2006); 
JOH07 --- Johnson et al. (2007); 
KJA82 --- Kj{\ae}rgaard et al. (1982); 
LAI07 --- Lai et al. (2007); 
LAI08 --- Lai et al. (2008); 
MCW95 --- McWilliam et al.(1995); 
MEL01 --- Mel\'{e}ndez, Barbuy, and Spite (2001); 
MEL02 --- Mel\'{e}ndez and Barbuy (2002); 
NOR97 --- Norris, Ryan, and Beers (1997); 
RAM11 --- Ram\'{\i}rez and Allende Prieto (2011); 
SIM04 --- Simmerer et al. (2004); 
SPI05 --- Spite et al. (2005); 
SPI06 --- Spite et al. (2006); 
TAK05 --- Takeda and Honda (2005);
TOM92 --- Tomkin et al. (1992);
ZHA11 --- Zhang et al. (2011). 
}}}
\endlastfoot
\hline
BD~+44~493   & {\it 5510} &  {\it 3.70}&  {\it 1.30}& {\it $-$3.68}& {\it $-$2.85}&({\it this study}) & {\it C I 10683/10685/10691} \\
            & 5510 &  3.70&  1.30& $-$3.68& $-$2.37& ITO09  & CH (4308--4316) \\
\hline
HD~126587    & {\it 4700} &  {\it 1.05}&  {\it 1.70}& {\it $-$3.16}& {\it $-$2.87}& ({\it this study}) & {\it C I 10683/10685/10691} \\
            & 4960 &  2.10&  1.80& $-$2.90& $-$2.59& HON04  & CH (4323) \\
            & 4910 &  1.85&  2.03& $-$2.84& $-$2.71& MCW95  & CH ($\sim$4300) \\
\hline
G~64-37 & {\it 6432} &  {\it 4.24}&  {\it 1.50}& {\it $-$3.08}& {\it $-$2.90}& ({\it this study}) & {\it C I 10683/10685/10691} [{\it 2009 data}] \\
        & {\it 6432} &  {\it 4.24}&  {\it 1.50}& {\it $-$3.08}& {\it $-$2.97}& ({\it this study}) & {\it C I 10683/10685/10691} [{\it 2011 data}] \\
            & 6432 &  4.24&  1.50& $-$3.08& $-$3.21& FAB09  & C I 9061/9062/9078/9088/9094/9111 \\
            & 6318 &  4.16&  1.50& $-$3.12& $-$2.83& AKE04  & C I 9061/9078/9094/9111 \\
\hline
BD~$-$18~5550  & {\it 4750} &  {\it 1.40}&  {\it 1.80}& {\it $-$3.06}& {\it $-$2.91}& ({\it this study}) & {\it C I 10683/10685/10691} \\
            & 4750 &  1.40&  1.80& $-$3.06& $-$3.08& CAY04  & CH (4224) \\
            & 4750 &  1.40&  1.80& $-$3.06& $-$3.08& SPI05  & CH (4224) \\
            & 4790 &  1.15&  2.14& $-$2.91& $-$2.91& MCW95  & CH ($\sim$4300) \\
            & 4683 &  1.70&  1.70& $-$2.87& $-$3.26& MEL02  & CH (4310) \\
            & 4750 &  1.40&  1.80& $-$3.06& $-$3.08& SPI06  & CH (4230) \\
            & 4806 &  1.72&  1.91& $-$2.89& $-$3.09& JOH07  & CH ($\sim$4300) \\
\hline
HD~115444    & {\it 4721} &  {\it 1.74}&  {\it 2.00}& {\it $-$2.71}& {\it $-$2.90}& ({\it this study}) & {\it C I 10683/10685/10691} \\
            & 4720 &  1.50&  1.70& $-$2.85& $-$3.26& HON04  & CH (4323) \\
            & 4775 &  1.68&  2.00& $-$2.86& $-$3.01& LAI07  & CH ($\sim$4300) \\
            & 4721 &  1.74&  2.00& $-$2.71& $-$3.45& SIM04  & CH (4315) \\
\hline
HD~122563    & {\it 4572} &  {\it 1.36}&  {\it 2.90}& {\it $-$2.61}& {\it $-$2.74}& ({\it this study}) & {\it C I 10683/10685/10691} \\
            & 4600 &  1.10&  2.00& $-$2.85& $-$3.22& CAY04  & CH (4224) \\
            & 4570 &  1.10&  2.20& $-$2.77& $-$3.18& HON04  & CH (4323) \\
            & 4600 &  1.10&  2.00& $-$2.82& $-$3.29& SPI05  & CH (4224) \\
            & 4650 &  1.40&  2.60& $-$2.68& $-$3.13& NOR97  & CH (4323) \\
            & 4600 &  1.10&  2.20& $-$2.62& $-$3.02& AOK05  & CH (4323) \\
            & 4600 &  1.10&  2.00& $-$2.82& $-$3.21& SPI06  & CH (4230) \\
            & 4615 &  1.27&  2.10& $-$2.47& $-$3.42& JOH07  & CH ($\sim$4300) \\
            & 4600 &  1.10&  2.20& $-$2.58& $-$2.94& AOK07  & CH (4323) \\
            & 4610 &  1.32&  2.00& $-$2.54& $-$2.94& LAI07  & CH ($\sim$4300) \\
            & 4572 &  1.36&  2.90& $-$2.61& $-$3.32& SIM04  & CH (4315) \\
\hline
BD~+23~3130  & {\it 5000} &  {\it 2.20}&  {\it 1.40}& {\it $-$2.60}& {\it $-$2.55}& ({\it this study}) & {\it C I 10683/10685/10691} \\
            & 5224 &  2.82&  2.00& $-$2.59& $-$2.29& LAI07  & CH ($\sim$4300) \\
            & 5285 &  2.83&  1.60& $-$2.62& $-$2.47& LAI08  & CH ($\sim$4300) \\
\hline
HD~140283    & {\it 5830} &  {\it 3.67}&  {\it 1.90}& {\it $-$2.55}& {\it $-$2.38}& ({\it this study}) & {\it C I 10683/10685/10691} \\
            & 5849 &  3.72&  1.50& $-$2.38& $-$2.48& FAB09  & C I 9061/9062/9078/9088/9094/9111 \\
            & 5630 &  3.50&  1.40& $-$2.53& $-$2.25& HON04  & CH (4323) \\
            & 5750 &  3.30&  1.20& $-$2.47& $-$2.31& AOK02  & CH (4323) \\
            & 5690 &  3.69&  1.50& $-$2.42& $-$2.21& AKE04  & C I 9061/9078/9094/9111 \\
            & 5640 &  3.28&  1.50& $-$2.64& $-$2.42& TOM92  & C I 9078/9088/9094/9111 \\
            & 5640 &  3.28&  1.50& $-$2.64& $-$2.40& TOM92  & CH ($\sim$4300) \\
\hline
G~186-26     & {\it 6417} &  {\it 4.42}&  {\it 1.50}& {\it $-$2.54}& {\it $-$2.43}& ({\it this study}) & {\it C I 10683/10685/10691} \\
            & 6417 &  4.42&  1.50& $-$2.54& $-$2.65& FAB09  & C I 9061/9062/9078/9088/9094/9111 \\
            & 6273 &  4.25&  1.50& $-$2.62& $-$2.41& AKE04  & C I 9061/9078/9094/9111 \\
\hline
HD~186478    & {\it 4730} &  {\it 1.50}&  {\it 1.80}& {\it $-$2.42}& {\it $-$2.70}& ({\it this study}) & {\it C I 10683/10685/10691} \\
            & 4700 &  1.30&  2.00& $-$2.60& $-$2.92& CAY04  & CH (4224) \\
            & 4720 &  1.60&  2.20& $-$2.50& $-$2.75& HON04  & CH (4323) \\
            & 4700 &  1.30&  2.00& $-$2.59& $-$2.89& SPI05  & CH (4224) \\
            & 4650 &  0.95&  2.71& $-$2.58& $-$2.86& MCW95  & CH ($\sim$4300) \\
            & 4700 &  1.30&  2.00& $-$2.59& $-$2.81& SPI06  & CH (4230) \\
            & 4831 &  1.78&  1.89& $-$2.63& $-$2.83& JOH07  & CH ($\sim$4300) \\
            & 4598 &  1.43&  2.00& $-$2.44& $-$2.96& SIM04  & CH (4315) \\
\hline
HD~108317    & {\it 5310} &  {\it 2.77}&  {\it 1.90}& {\it $-$2.35}& {\it $-$2.22}& ({\it this study}) & {\it C I 10683/10685/10691} \\
            & 5234 &  2.68&  2.00& $-$2.28& $-$2.23& SIM04  & CH (4315) \\
\hline
BD~+26~2606  & {\it 5875} &  {\it 4.10}&  {\it 0.40}& {\it $-$2.30}& {\it $-$2.07}& ({\it this study}) & {\it C I 10683/10685/10691} \\
            & 5910 &  3.63&  1.50& $-$2.63& $-$2.35& TOM92  & C I 9078/9088/9094/9111 \\
            & 5910 &  3.63&  1.50& $-$2.63& $-$2.40& TOM92  & CH ($\sim$4300) \\
\hline
HD~6268      & {\it 4735} &  {\it 1.61}&  {\it 2.10}& {\it $-$2.30}& {\it $-$2.91}& ({\it this study}) & {\it C I 10683/10685/10691} \\
            & 4600 &  1.00&  2.10& $-$2.63& $-$3.30& HON04  & CH (4323) \\
            & 4670 &  0.75&  2.73& $-$2.59& $-$3.26& MCW95  & CH ($\sim$4300) \\
            & 4705 &  1.50&  1.90& $-$2.35& $-$3.48& MEL02  & CH (4310) \\
\hline
HD~13979     & {\it 5075} &  {\it 1.90}&  {\it 1.30}& {\it $-$2.26}& {\it $-$3.20}& ({\it this study}) & {\it C I 10683/10685/10691} \\
            & 4970 &  1.15&  2.35& $-$2.65& $-$3.25& MCW95  & CH ($\sim$4300) \\
\hline
HD~19445     & {\it 6130} &  {\it 4.39}&  {\it 2.10}& {\it $-$2.05}& {\it $-$2.01}& ({\it this study}) & {\it C I 10683/10685/10691} \\
            & 6047 &  4.51&  0.80& $-$1.94& $-$1.99& GRA00  & CH (4235, 4365) \\
            & 5880 &  4.40&  1.50& $-$2.15& $-$1.76& TOM92  & C I 9078/9088/9094/9111 \\
            & 5880 &  4.40&  1.50& $-$2.15& $-$2.20& TOM92  & CH ($\sim$4300) \\
\hline
HD~221170    & {\it 4560} &  {\it 1.37}&  {\it 1.60}& {\it $-$2.00}& {\it $-$2.44}& ({\it this study}) & {\it C I 10683/10685/10691} \\
            & 4648 &  1.57&  2.22& $-$2.15& $-$2.70& BAR05  & CH (4310--4313, 4362--4367) \\
            & 4460 &  0.75&  1.60& $-$2.11& $-$2.41& MEL01  & CH ($\sim$4300) [literature value adopted] \\
            & 4510 &  1.00&  1.80& $-$2.20& $-$2.89& IVA06  & CH ($\sim$4300) \\
            & 4410 &  1.09&  1.70& $-$2.03& $-$2.85& SIM04  & CH (4315) \\
            & 4648 &  1.57&  2.22& $-$2.14& $-$2.71& ZHA11  & CH (4310) \\
            & 4600 &  1.50&  1.90& $-$1.98& $-$2.59& AOK08  & CH (4323) \\
\hline
HD~216143    & {\it 4525} &  {\it 1.77}&  {\it 1.90}& {\it $-$1.92}& {\it $-$2.08}& ({\it this study}) & {\it C I 10683/10685/10691} \\
            & 4360 &  0.50&  1.80& $-$2.22& $-$2.87& MEL01  & CO (1.56$\mu$m, 2.33$\mu$m) \\
            & 4450 &  0.80&  2.30& $-$2.27& $-$2.64& AOK08  & CH (4323) \\
\hline
HD~204543    & {\it 4672} &  {\it 1.49}&  {\it 2.00}& {\it $-$1.72}& {\it $-$2.15}& ({\it this study}) & {\it C I 10683/10685/10691} \\
            & 4570 &  1.24&  2.00& $-$1.98& $-$2.43& LAI07  & CH ($\sim$4300) \\
            & 4672 &  1.49&  2.00& $-$1.72& $-$2.42& SIM04  & CH (4315) \\
            & 4600 &  1.00&  2.20& $-$1.81& $-$2.44& AOK08  & CH (4323) \\
\hline
G~29-23      & {\it 6194} &  {\it 4.04}&  {\it 1.50}& {\it $-$1.69}& {\it $-$1.62}& ({\it this study}) & {\it C I 10683/10685/10691} \\
            & 6194 &  4.04&  1.50& $-$1.69& $-$1.91& FAB09  & C I 9061/9062/9078/9088/9094/9111 \\
            & 5966 &  3.82&  1.50& $-$1.80& $-$1.70& AKE04  & C I 9061/9078/9094/9111 \\
\hline
G~18-39      & {\it 6093} &  {\it 4.19}&  {\it 1.50}& {\it $-$1.46}& {\it $-$1.40}& ({\it this study}) & {\it C I 10683/10685/10691} \\
            & 6093 &  4.19&  1.50& $-$1.46& $-$1.45& FAB09  & C I 9061/9062/9078/9088/9094/9111 \\
            & 5910 &  4.09&  1.50& $-$1.52& $-$1.26& AKE04  & C I 9061/9078/9094/9111 \\
\hline
HD~219617NE  & {\it 5825} &  {\it 4.30}&  {\it 1.40}& {\it $-$1.40}& {\it $-$1.41}& ({\it this study}) & {\it C I 10683/10685/10691} \\
HD~219617SW  & {\it 5825} &  {\it 4.30}&  {\it 1.40}& {\it $-$1.40}& {\it $-$1.44}& ({\it this study}) & {\it C I 10683/10685/10691} \\
            & 5974 &  3.88&  1.23& $-$1.48& $-$1.49& GRA00  & CH (4235, 4365) \\
\hline
HD~121135    & {\it 4934} &  {\it 1.91}&  {\it 1.60}& {\it $-$1.37}& {\it $-$1.42}& ({\it this study}) & {\it C I 10683/10685/10691} \\
            & 4934 &  1.91&  1.60& $-$1.37& $-$1.99& SIM04  & CH (4315) \\
\hline
HD~166161    & {\it 5350} &  {\it 2.56}&  {\it 2.25}& {\it $-$1.22}& {\it $-$1.12}& ({\it this study}) & {\it C I 10683/10685/10691} \\
            & 5270 &  2.51&  1.55& $-$1.08& $-$1.58& GRA00  & CH (4235, 4365) \\
            & 5350 &  2.56&  2.25& $-$1.22& $-$1.33& SIM04  & CH (4315) \\
\hline
HD~194598    & {\it 6020} &  {\it 4.30}&  {\it 1.40}& {\it $-$1.15}& {\it $-$1.19}& ({\it this study}) & {\it C I 10683/10685/10691} \\
            & 6046 &  4.31&  0.80& $-$1.08& $-$1.08& GRA00  & CH (4235, 4365) \\
            & 5906 &  4.25&  1.30& $-$1.17& $-$1.07& AKE04  & C I 9061/9078/9094/9111 \\
            & 6044 &  4.19&  1.00& $-$1.16& $-$1.13& SIM04  & CH (4315) \\
            & 5910 &  4.28&  1.50& $-$1.32& $-$1.12& TOM92  & C I 9078/9088/9094/9111 \\
            & 5910 &  4.28&  1.50& $-$1.32& $-$1.50& TOM92  & CH ($\sim$4300) \\
\hline
HD~193901    & {\it 5699} &  {\it 4.42}&  {\it 1.20}& {\it $-$1.10}& {\it $-$1.15}& ({\it this study}) & {\it C I 10683/10685/10691} \\
            & 5823 &  4.58&  0.80& $-$1.10& $-$1.21& GRA00  & CH (4235, 4365) \\
            & 5672 &  4.38&  1.00& $-$1.12& $-$1.14& AKE04  & C I 9061/9078/9094/9111 \\
            & 5750 &  4.46&  1.50& $-$1.16& $-$1.28& SIM04  & CH (4315) \\
            & 5810 &  4.83&  1.50& $-$1.22& $-$1.26& TOM92  & C I 9078/9088/9094/9111 \\
            & 5810 &  4.83&  1.50& $-$1.22& $-$1.90& TOM92  & CH ($\sim$4300) \\
\hline
HD~201891    & {\it 5900} &  {\it 4.19}&  {\it 1.40}& {\it $-$1.10}& {\it $-$1.01}& ({\it this study}) & {\it C I 10683/10685/10691} \\
            & 5991 &  4.30&  1.10& $-$1.13& $-$0.92& GRA00  & CH (4235, 4365) \\
            & 5909 &  4.19&  1.00& $-$1.10& $-$1.04& SIM04  & CH (4315) \\
\hline
HD~148816    & {\it 5860} &  {\it 4.07}&  {\it 1.60}& {\it $-$1.00}& {\it $-$0.73}& ({\it this study}) & {\it C I 10683/10685/10691} \\
            & 5823 &  4.14&  1.20& $-$0.73& $-$0.53& AKE04  & C I 9061/9078/9094/9111 \\
\hline
HD~6582      & {\it 5331} &  {\it 4.54}&  {\it 0.73}& {\it $-$0.81}& {\it $-$0.66}& ({\it this study}) & {\it C I 10683/10685/10691} \\
            & 5408 &  4.40&  0.80& $-$0.83& $-$0.79& MEL01  & CO (1.56$\mu$m, 2.33$\mu$m) \\
\hline
Arcturus    & {\it 4281} &  {\it 1.72}&  {\it 1.49}& {\it $-$0.55}& {\it $-$0.39}& ({\it this study}) & {\it C I 10683/10685/10691} \\
            & 4350 &  1.80&  1.70& $-$0.51& $-$0.67& KJA82  & C$_{2}$ (5630) \\
            & 4286 &  1.66&  1.74& $-$0.52& $-$0.09& RAM11  & C I 5380/8335/9078/9111 \\
\hline
HD~142373    & {\it 5776} &  {\it 3.83}&  {\it 1.26}& {\it $-$0.51}& {\it $-$0.55}& ({\it this study}) & {\it C I 10683/10685/10691} \\
            & 5776 &  3.83&  1.26& $-$0.51& $-$0.38& TAK05  & C I 5052/5380 \\
\hline
HD~165908    & {\it 6183} &  {\it 4.35}&  {\it 1.24}& {\it $-$0.46}& {\it $-$0.58}& ({\it this study}) & {\it C I 10683/10685/10691} \\
            & 6183 &  4.35&  1.24& $-$0.46& $-$0.24& TAK05  & C I 5052/5380 \\
\hline
HD~10700     & {\it 5420} &  {\it 4.68}&  {\it 0.66}& {\it $-$0.43}& {\it $-$0.42}& ({\it this study}) & {\it C I 10683/10685/10691} \\
            & 5420 &  4.68&  0.66& $-$0.43& $-$0.20& TAK05  & C I 5052/5380 \\
\hline
HD~131156    & {\it 5527} &  {\it 4.60}&  {\it 1.10}& {\it $-$0.13}& {\it $-$0.13}& ({\it this study}) & {\it C I 10683/10685/10691} \\
            & 5527 &  4.60&  1.10& $-$0.13& $-$0.14& TAK05  & C I 5052/5380 \\
\hline
HD~141004    & {\it 5877} &  {\it 4.11}&  {\it 1.17}& {\it $-$0.01}& {\it $-$0.05}& ({\it this study}) & {\it C I 10683/10685/10691} \\
            & 5877 &  4.11&  1.17& $-$0.01&  0.02& TAK05  & C I 5052/5380 \\
\hline
HD~196755    & {\it 5750} &  {\it 3.83}&  {\it 1.23}&  {\it 0.09}& {\it $-$0.09}& ({\it this study}) & {\it C I 10683/10685/10691} \\
            & 5750 &  3.83&  1.23&  0.09& $-$0.05& TAK05  & C I 5052/5380 \\
\hline
HD~161797    & {\it 5580} &  {\it 3.99}&  {\it 1.11}&  {\it 0.29}&  {\it 0.25}& ({\it this study}) & {\it C I 10683/10685/10691} \\
            & 5580 &  3.99&  1.11&  0.29&  0.27& TAK05  & C I 5052/5380 \\
\hline
HD~182572    & {\it 5566} &  {\it 4.11}&  {\it 1.07}&  {\it 0.33}&  {\it 0.35}& ({\it this study}) & {\it C I 10683/10685/10691} \\
            & 5566 &  4.11&  1.07&  0.33&  0.39& TAK05  & C I 5052/5380 \\
\hline
%\end{tabular}
%\end{center}
%\end{table}
\end{longtable}

\end{document}